\documentclass[12pt]{article}
\usepackage{amsmath}
\usepackage{amsthm}
\usepackage{amssymb}
\usepackage{amsfonts}
\usepackage{epic}
\usepackage{eepic}
 \usepackage{times}
\usepackage[matrix,arrow]{xy}

\usepackage{macros}

\usepackage{epsfig}

\theoremstyle{plain}

\theoremstyle{remark}

\newcommand{\sst}{}

\newcommand{\nn}{\nonumber}
\newcommand{\bbra}{\bra\!\bra}
\newcommand{\kket}{\ket\!\ket}

\newcommand{\lket}{\ket_{\rm\sst L}^{}}

\newcommand{\xket}{\ket_{\rm\sst X_0}^{}}

\newcommand{\xbra}[1]{\bra \, #1 \,|{}_{\rm\sst X_0}^{}}
\renewcommand{\1}{\one}
\renewcommand{\2}{\two}
\newcommand{\3}{\three}

\newcommand{\beq}{\begin{equation}}
\newcommand{\eeq}{\end{equation}}

\newcommand{\pa}{\partial}
\newcommand{\ot}{\otimes}
\newcommand{\ra}{\to}

\newcommand{\bra}{\langle}

\newcommand{\ket}{\rangle}

\newcommand{\al}{\alpha}

\newcommand{\be}{\beta}

\newcommand{\Ga}{\Gamma}
\newcommand{\de}{\delta}
\newcommand{\De}{\Delta}
\newcommand{\ep}{\epsilon}
\newcommand{\la}{\lambda}
\newcommand{\om}{\omega}
\newcommand{\Om}{\Omega}
\newcommand{\si}{\sigma}

\newcommand{\bz}{\bar{z}}

\newcommand{\CC}{{\mathcal C}}

\newcommand{\CH}{{\mathcal H}}

\newcommand{\CO}{{\mathcal O}}
\newcommand{\CP}{{\mathcal P}}

\newcommand{\CV}{{\mathcal V}}

\newcommand{\mub}{\mu_{\rm B}}

\newcommand{\one}{{\mathfrak 1}}
\newcommand{\two}{{\mathfrak 2}}
\newcommand{\three}{{\mathfrak 3}}

\newcommand{\BR}{{\mathbb R}}

\newcommand{\BU}{{\mathbb U}}

\newcommand{\BS}{{\mathbb S}}

\DeclareMathOperator{\sgn}{sgn}
\DeclareMathOperator*{\Res}{Res}

\newcommand{\dasym}{\underset{\de\ra 0}{\sim}}

\newcommand{\rf}[1]{(\ref{#1})}

\newcommand{\aufz}
{\begin{list}{$\bullet$}{\topsep0cm \itemsep0cm \parsep0cm}}
\newcommand{\eaufz}{\end{list}}
\begin{document}
\title{On the
time-dependent description for the decay of unstable D-branes}
\author{K,Graham${}^{\1}$, A. Konechny${}^{\2}$ and J. Teschner${}^{\3}$}
\address{${}^\1$ Institut f\"ur theoretische Physik, Freie Universit\"at Berlin,
Arnimallee 14, 14195 Berlin, Germany\\[1ex]
${}^\2$ Department of Physics and Astronomy, Rutgers, The State University of New Jersey\\
Piscataway, NJ 08854, USA\\[1ex]
${}^\3$ DESY Theory,
Notkestr. 85
22607 Hamburg
Germany}
\maketitle
\begin{quote}
{\small We discuss how to describe
time-dependent phenomena in string theory like
the decay of unstable D-branes with the help
of the world-sheet formulation. It is shown in a
nontrivial well-controlled example that
the coupling of the tachyons to propagating
on-shell modes which escape to infinity
can lead to time-dependent relaxation into a
stationary final state. The final state corresponds
to a fixed point of the RG flow generated by
the relevant field from which the tachyon vertex
operator is constructed. On the way we set up a
fairly general formalism for the description
of slow time-dependent phenomena with the help of
conformal perturbation theory on the world-sheet.}
\end{quote}

\section{Introduction}

Time-dependent phenomena in string theory are not easy to understand.
Tractable examples of time-dependent backgrounds are rare, making it hard to
extract general lessons about how string theory differs from ordinary field
theory in the description of time-dependent phenomena. Conceptual and
technical issues appear to be intertwined in a complicated way:
If one tries to construct the conformal field theories (CFTs) describing
time-dependent backgrounds using nonlinear sigma  models,
one has to face the unboundedness of the sigma model path
integral due to the indefinite signature of the metric.
The very definition of CFTs describing time-dependent backgrounds
is therefore problematic in general.
On the other hand, experience
from point-particle field theories suggests
the definition of the vacuum of the string (field) theory becomes
ambiguous in generic time-dependent backgrounds. If and how such technical
and conceptual problems are intertwined is not well understood.

An interesting class of time-dependent phenomena is often referred to as
tachyon condensation. This includes decay processes of unstable
D-branes, interpreted as the condensation of an
open string tachyon on the relevant brane. In order to bypass the
technical difficulties involved in the definition of
a CFT which describes a tachyon condensation process it is
often proposed that one can simply restrict attention to the
CFT describing the spatial part of the background. Tachyons
are then described by relevant fields in this CFT. Perturbing the
CFT by these relevant fields will generate nontrivial renormalization group (RG) flows,
whose end-points are then conjecturally interpreted as the
possible final states of decay processes in the genuinely
time-dependent description.

The scope and limitations of this picture are not well understood. A major
puzzle stems from the fact that the evolution equations for background
fields like the tachyon are second order in time
derivatives, whereas the RG flow equations are first order
differential equations. This seems to exclude
any simple relationship between renormalization group flows
and the true time evolution.
One may still hope the
time-independent description using RG flows correctly
describes at least certain qualitative features of the true time evolution
like the initial and final states.
However, the fact that the time evolution equations are second order
differential equations suggests that
 the solutions which describe rolling from an
unstable maximum of the tachyon potential will typically
exhibit oscillatory behavior and may not relax
to any stationary final state as suggested intuitively
by the time-independent picture in terms of
RG flows.


A first step towards the resolution of this puzzle was recently
made in \cite{FHL}. Proper inclusion of the dilaton typically
produces a damping force that may lead to relaxation into the
minimum of the tachyon potential which corresponds to the
end-point of an RG flow in the time-independent description.
However, the mechanism studied in \cite{FHL}
applies only to supercritical backgrounds.

In the present paper we will propose a different mechanism
which reconciles - qualitatively, for a certain class of backgrounds -
the time-dependent and the time-independent pictures.
The mechanism applies to backgrounds in which a localized tachyonic
excitation couples to a continuum of string modes which can escape to
infinity.
The energy released in the condensation of the localized
tachyon is radiated away to infinity. Having infinitely many
degrees of freedom into which the energy is dissipated avoids
any oscillatory nature for the resulting dynamics of the tachyon.

The noncompactness of the string background is crucial for this mechanism.
If the (open) string spectrum is discrete, as is typically the case for
compact backgrounds (branes),
it may not be true that the system relaxes into a new
stationary state.

\subsubsection{On the use of conformal perturbation theory.}

It is not trivial to construct examples which illustrate the points
above within the world-sheet approach to string theory.
The examples we will discuss in this paper will be constructed
by means of conformal perturbation theory. However, the application of conformal
perturbation theory turns out to be somewhat subtle for this type
of problem.

Conformal perturbation theory will
be useful in the present
context only if there is a parameter $\de$ in the
theory which allows one to make the decay process arbitrarily slow.
The parameter $\de$ is related to the deviation from marginality
in the time-independent picture. It also controls the
amount of energy stored in the unstable brane.
One ultimately aims to develop a series expansion in the parameter $\de$.

In some cases we will be dealing with examples where, strictly speaking, no
renormalization of the perturbation Lagrangian is necessary.
This means the bare coupling constant $\la$ is well-defined,
and the series expansion in $\la$ is meaningful. However, the
expansion in $\la$ is not useful to
extract the corrections to leading order in $\de$.
The use of renormalization group (RG) technology will
prove crucial in order to extract these contributions to string emission
amplitudes. We will take advantage of the fact
that using effective
coupling constants as determined by the renormalization group equations
effectively amounts to resumming certain contributions to
a ``naive'' perturbative expansion.
The fact itself is certainly known from other quantum field theoretical
models, but crucial for our present investigation
is a more precise statement: Any ``proper'' regularization
scheme\footnote{See Section \ref{summingRG} for the explanation of
what we call a proper regularization scheme.}
leads to the definition of renormalized coupling constants $\la_{\rm ren}$
which themselves are of the order $\de$, and which allow
us to capture the leading order (in $\de$) effects in the
first order of perturbation theory in $\la_{\rm ren}$.
This result may be known, but at least
in the context of conformal perturbation theory we did
not find a sufficiently detailed discussion of it
in the literature
\footnote{See, however, \cite{CL} for some remarks
in this direction}. We have therefore
included a self-contained
discussion of this point in our paper.

A second subtlety is the following.
If one uses conformal perturbation theory one might be tempted to
drop irrelevant fields from the perturbation Lagrangian.
This turns out to lead to incorrect results in our example.
Irrelevant fields that are sufficiently close to
marginality produce important contributions to
correlation functions. In our case they precisely take care of the
radiation into strings that propagate out to infinity.

\subsubsection{The model.}

The model which will illustrate the mechanism
proposed above in a controlled way is the so-called $c=1$ noncritical string
theory, see e.g. \cite{GM,SS} for reviews.
The $c=1$ string theory is a two-dimensional string background
with coordinates $(X_0,\phi)\in\BR^2$, where $X_0$ represents
time. This background is characterized by the following
expectation values for the target-space metric $G_{\mu\nu}$,  the
dilaton $\Phi$ and the tachyon field $T$:
\begin{equation}\label{backgr}
G_{\mu\nu}=\eta_{\mu\nu},\quad \Phi=\phi,\quad T=\mu e^{2\phi}\,.
\end{equation}
The worldsheet-description of this background  
is defined by a (boundary) CFT with central
charge $c=26$ which is the product of Liouville theory
with the CFT of a free timelike boson.

Note the linear growth of the dilaton in \rf{backgr} implies
the string coupling is strong for $\phi\ra\infty$, whereas it is weak for
$\phi\ra-\infty$. The tachyon expectation value $T=\mu e^{2\phi}$
produces a force on the closed strings which exponentially grows
for $\phi\ra\infty$, and which therefore effectively confines the
closed strings to the weak coupling region $\phi\ra-\infty$. This force
implies that the closed string states which for early times
are injected into the weak coupling region with positive momentum will
ultimately be reflected back into the weak coupling region.

An open string sector can be introduced by imposing Neumann-type
boundary conditions for both the space and the time directions
that may include a space-dependent force on
the end-points of the open strings.
The corresponding D-branes are often called FZZT branes.
These branes are described by a single parameter $\de$ which
may be thought of as a parametrization for the force on the
end points of the open strings. It should be emphasized, however,
that we are dealing with a case in which semiclassical
reasoning is not applicable, and where the
stringy corrections are substantial. What can be deduced from
the exact solution of boundary Liouville theory
\cite{FZZ,T1,Ho,PT,T2} is the following:
The open string spectrum always contains propagating open strings
which bounce off the potential wall coming from the tachyon
condensate $T=\mu e^{2\phi}$.
However, for certain values of the parameter $\de$ the string spectrum
also contains bound states.
In the case considered in this paper there is a
single bound state $|\vartheta\kket$ which turns out to be tachyonic.
The wave-function
of $|\vartheta\kket$ decays exponentially towards the weak
coupling region. This means the FZZT
brane is carrying a {\it localized} open string
tachyon.

It is natural to interpret
the bound state as the result of a balance between
an attractive force on the end-points of the strings and
the repulsive force acting on the bulk of the string.
The very existence of the bound state is an
indication of the presence of a sink in some effective potential
for open  strings on the FZZT branes.
We expect the condensation of the open string
tachyon $|\vartheta\kket$ may ``fill up'' this sink,
producing an FZZT brane on which the forces on the open strings
are effectively repulsive.
The energy released in this process
will be radiated away into the
weak coupling region as open and
closed strings. We will show this is precisely
what happens by constructing a time-dependent solution
to noncritical string theory describing tachyon condensation
on FZZT branes. One should note that the
effect of closed string radiation will be subleading
in the string coupling constant $g_s$ in the model we will
study.


\subsubsection{Content of the paper}

The main technical result of the paper is the perturbative
construction of a time-dependent solution of noncritical
open-closed string theory to lowest order in $\de$.
This will heavily exploit the RG improvement to the
perturbative expansion. We will therefore start in section
\ref{summingRG} by explaining how RG techniques
manage to resum the contributions to leading order in $\de$.

In the following section we discuss the
construction of general time-dependent backgrounds describing
open string tachyons.  In so doing we will see some of the
limitations of time-independent RG flows as a description of
time-dependent phenomena.

Section \ref{LBCFT} introduces the
relevant aspects of boundary
Liouville theory, somewhat sharpening the physical picture of the D1
branes along the way.
We also discuss a renormalization group flow obtained
by perturbing Liouville theory with a relevant
boundary field that provides a
first hint towards the scenario discussed in the introduction. The end-point
of the RG flow which starts from the boundary condition
$\de$ is identified with the boundary condition associated
to the parameter $-\de$.

Section \ref{background} then discusses the
perturbative construction of a time-dependent
solution of string theory which is associated to the
condensation of the open string tachyon on the D1
branes. It is found that
the time dependent solution smoothly interpolates
between the static solution with parameter $+\delta$
in the asymptotic past
and a background which can be seen as the
static solution with parameter $-\delta$ plus a propagating radiation
background in the infinite future.

Section 6 contains a discussion of the  results obtained in the paper and
directions for future work.

Appendix A discusses RG  equations in the presence of  UV divergences  which appear
 when the identity operator is present in the OPE's of perturbing operators.
 Appendix B contains technical details regarding the boundary Liouville theory.

\section{Renormalization group improvement of perturbative series}
\label{summingRG}
\setcounter{equation}{0}

To begin with, we will consider the perturbation  of a given boundary  conformal field theory (BCFT)
by a relevant boundary field with conformal dimension $1-\de$.
Our aim is to show how renormalisation group (RG)
technology provides an
efficient tool to extract the leading behaviour of correlation functions for $\de\ra 0$.

To explain the method, we will first consider cases in which all short distance
singularities are integrable, so no renormalization is necessary.
RG technology is nevertheless found to be an efficient tool for
reaching our aim. The existence of divergencies in the
perturbative integrals does not change any relevant
feature of the resulting picture, as shown in appendix A.

We will consider a
BCFT which has a family of boundary
conditions $B_\de$ parametrized by a parameter $\de$.
The BCFT under consideration will contain
primary boundary fields $\phi_i({\tau})$ of conformal weight $h_i$,
living on the unit circle, $0\le\tau<2\pi$,
The labels $i$ take values in some set $F$ (in
the examples to follow, $F$ may be a continuum).

To
apply perturbation theory usefully, we will assume that the boundary
conditions $B_\de$ are parametrized by $\de$ in such a way that
a small value for $\de$ implies a relevant boundary
field $\phi_0$ is nearly marginal in the sense that
$y_0\equiv 1{-}h_0=\CO(\de)$.  It will also be important to consider
all the other nearly marginal
fields, including those which are irrelevant.
We will denote this set by $M \subset F$, the elements $\phi_i(\tau)$
of $M$ satisfy  $y_i = 1{-}h_i = \CO(\delta)$.
For simplicity we assume all other fields have $y_i = 1{-}h_i = \CO(1)$.%
\footnote{These assumptions, together with a later one concerning the OPE coefficients, can be weakened or adjusted.  In fact, in our string theory example $y_i = \CO( \delta^2)$ and $C_{ij}{}^k= \CO(\delta)$.  It takes a moment to see such changes do not affect the conclusions.}

The correlation functions of fields
$\phi_i(\tau)$ can be evaluated using their OPE, valid
for $0<\tau<\pi$, 
\begin{equation}\label{eq:OPE}
   \phi_i(\tau)\phi_j(0) = \sum_{k \in F} C_{ij}{}^k \left( 2 \sin{ \tfrac \tau 2 } \right)^{-1+y_i+y_j-y_k}
  \phi_k(0) + \text{descendents} \; .
\end{equation}
Note that if $F$ contains a continuum, the sum will become an integral.
For simplicity, we also assume the OPE coefficients have a nice $\delta$-dependence like $C_{ij}{}^k = \CO(1)$.

We will consider a perturbation of the boundary condition by an
operator 
$\phi_0(\tau)$, $0 \in M$,
\begin{equation}
  S_\text{pert} = \lambda \int_{\partial \Sigma} d\tau \phi_0(\tau) \; ,
\end{equation}
which defines perturbed correlation functions via
\begin{align}\label{eq:pertdefn}
  \langle \ldots \rangle_\lambda
&   = \langle \ldots e^{-S_\text{pert}} \rangle^{}_{B_{\de}}\\
&  = \langle \ldots  \rangle^{}_{B_{\de}} - \lambda
\int\limits_{\partial \Sigma}  d\tau
   \langle \ldots \phi_0(\tau) \rangle^{}_{B_{\de}} + \frac{\lambda^2}{2}
   \int\limits_{\partial \Sigma \times \partial \Sigma}
 d\tau_1d\tau_2
   \langle \ldots \phi_0(\tau_1)\phi_0(\tau_2) \rangle^{}_{B_{\de}} + \ldots \; .
\nn\end{align}
In the examples to follow, this perturbation will be UV finite and so
the coupling constant $\lambda$ is well defined and the perturbation
series in $\lambda$ does not require any renormalisation.  Our task for
this section is to calculate the leading $\delta$ behaviour of this
series. We will find that RG technology provides the answer but to see
why, it helps to attempt to extract the leading $\delta$ behaviour by
brute force.

\subsection{Brute force calculation}

We concentrate on the perturbation expansion \eqref{eq:pertdefn} in the
particular case where dots denote the insertion of operators into the
interior of the disc.
We proceed by taking the $\la$ series \eqref{eq:pertdefn} and studying
the small $\de$ behaviour of each term.  The $n$th term will be found
to go as $\la^n \delta^{1-n}$.  If $\la=\CO(\de)$ it is clear the
leading correction is $\CO(\de)$, however every order makes a contribution.
Our task is to calculate these contributions.

Consider the second order term of \eqref{eq:pertdefn}.  As $\delta
\to 0$, this term becomes divergent due to the singularity in
the OPE \eqref{eq:OPE}.  To extract this singular behaviour, we introduce
a distance $L$ inside which the boundary-boundary OPE is valid (this
distance will depend on the location of the interior operators).  We
write,
\begin{equation}\label{eq:split2}
    \frac{\la^2}{2} \int\limits_{\pa \Sigma} d\tau_1
\int\limits_{\pa\Sigma}d\tau_2 \;\bra\dots \phi_0(\tau_1) \phi_0(\tau_2) \ket^{}_{B_{\de}}
  = \frac{\la^2}{2}\!\int\limits_{|\tau_1{-}\tau_2|<L} \!\!\!
d\tau_1d\tau_2 \ldots  + \frac{\la^2}{2}\!\int\limits_{|\tau_1{-}\tau_2|>L} \!\!\! d\tau_1d\tau_2 \ldots \; ,
\end{equation}
and the first term becomes,
\begin{align}
 \frac{\la^2}{2}\,\int_{|\tau_1-\tau_2|<L} \hspace{-25pt} d\tau_1d\tau_2 &
\bra\dots \phi_0(\tau_1) \phi_0(\tau_2) \ket^{}_{B_{\de}} \,\notag\\
 &= \la^2 \sum_{k \in F} C_{00}{}^k \int_0^L du \left( 2 \sin \tfrac u2 \right)^{-1 +2y_0 -y_k}
\int_{\pa \Sigma} \! d\tau \bra\dots \phi_k(\tau) \ket^{}_{B_{\de}} + \text{descendents} \notag \\
 &= \la^2 \sum_{k \in M} C_{00}{}^k \frac{1}{2y_0 -y_k}
\int_{\pa \Sigma} \! d\tau \bra\dots \phi_k(\tau) \ket^{}_{B_{\de}} + \text{subleading in $\delta$} \; .
\end{align}
The divergence is independent of our choice of $L$ and comes from the fusion of the perturbation into almost marginal fields (the fields with $y_i = \CO(\delta)$, both relevant and irrelevant).
The other fields contribute to a higher order in $\delta$.
Also note that the geometry of the boundary is largely immaterial for this calculation, only the short distance behaviour of the OPE is important.
Finally, second term in \eqref{eq:split2} is finite as $\delta \to 0$ and so represents a subleading contribution to the correlation function.

Moving to third order, the leading $\delta$ behaviour will come from the region of integration where all the boundary fields come together simultaneously.  We again introduce a distance $L$ to isolate this contribution and use the OPE to evaluate the integrals.  Note the OPE is only valid for 
$\tau>0$ so one must be careful:
\begin{align}
    -\frac{\la^3}{6}\, \int_{|\tau_i{-}\tau_j|<L} \hspace{-25pt} d\tau_1d\tau_2d\tau_3 &
\bra\dots \phi_0(\tau_1) \phi_0(\tau_2) \phi_0(\tau_3) \ket^{}_{B_{\de}}  \\
  &= -\la^3\, \int d\tau\int_0^{L}du_1\int_{u_1}^{L}du_2
\bra\dots \phi_0(\tau{+}u_2) \phi_0(\tau{+}u_1) \phi_0(\tau) \ket^{}_{B_{\de}} \notag \\
  &= -\la^3\, \int d\tau\int_{0}^{\tfrac 12 u_2}du_1\int_{0}^{L}du_2
 \ldots
 - \la^3\, \int d\tau\int_{\tfrac 12 u_2}^{u_2}du_1\int_{0}^{L}du_2
 \ldots \nn \; .
\end{align}
Applying the OPE in the first term,
\begin{align}
  -\la^3 \! \sum_{k,\ell \in S} \! & C_{00}{}^k  C_{0k}{}^\ell \!\!\int \!\! d\tau  \!\! \int_{0}^{\tfrac 12 u_2} \!\!\!\!\!\! du_1  \!\! \int_{0}^{L} \!\!\!\! du_2 \!
  \left( 2 \sin \tfrac{u_1}{2} \right)^{{-}1{+}2y_0{-}y_k} \! \left( 2 \sin \tfrac{u_2}{2} \right)^{{-}1{+}y_0{+}y_k{-}y_\ell} \!\! \bra\dots \phi_\ell(\tau) \ket^{}_{B_{\de}} \nn \\[-2ex] &
\hspace{10cm} {+} \text{descendants} \notag \\[-1ex]
  &= -\la^3\, \sum_{k,\ell \in M} \frac{ C_{00}{}^k C_{0k}{}^\ell }{ (2y_0 - y_k)(3y_0 - y_\ell )}
   \int d\tau \bra\dots \phi_\ell(\tau) \ket^{}_{B_{\de}} + \text{subleading in $\delta$} \; ,
\end{align}
while the second term gives,
\begin{align}
  -\la^3 \! \sum_{k,\ell \in S} \! &
C_{00}{}^k  C_{k0}{}^\ell \!\!\int \!\! d\tau  \!\! \int_{\tfrac 12 u_2}^{u_2} \!\!\!\!\!\! du_1  \!\! \int_{0}^{L} \!\!\!\! du_2 \!
  \left( 2 \sin \tfrac{u_2{-}u_1}{2} \right)^{{-}1{+}2y_0{-}y_k} \! \left( 2 \sin \tfrac{u_1}{2} \right)^{{-}1{+}y_0{+}y_k{-}y_\ell} \!\! \bra\dots \phi_\ell(\tau) \ket^{}_{B_{\de}} \nn\\[-2ex] & \hspace{10cm}  {+} \text{descendants} \notag \\[-1ex]
  &= -\la^3\, \sum_{k,\ell \in M} \frac{ C_{00}{}^k C_{k0}{}^\ell }{ (2y_0 - y_k)(3y_0 - y_\ell )}
   \int d\tau \bra\dots \phi_\ell(\tau) \ket^{}_{B_{\de}} + \text{subleading in $\delta$} \; .
\end{align}
Combining these two contributions gives the leading third order term.

At this point we repeat some important observations, also valid at higher orders in $\la$.  1) Only the almost marginal fields contribute, even in the intermediate stages.  Other fields have $y$-values $\CO(1)$ and so make subleading contributions.
2) Although this was on the disc, the result is largely independent of geometry.

One may repeat this analysis at each order in $\la$.  Leading $\delta$ behaviour will come from the region of integration where all the perturbing fields come together simultaneously, can be evaluated using successive OPEs and will behave as $\la^n \delta^{1-n}$.  The leading $\delta$ contribution of a general correlator is of the generic form,
\begin{gather}
    \langle \ldots \rangle_\lambda
   = \langle \ldots  \rangle^{}_{B_{\de}} - \sum_{k \in M} \hat{\la}_k \int_{\partial \Sigma} \!\!\!
 d\tau \langle \ldots \phi_k(\tau) \rangle^{}_{B_{\de}} + \text{subleading in $\delta$} \; ,
 \label{eq:ans1} \\
  \hat{\la}_0 = \la - \la^2 \frac{C_{00}{}^0}{y_0}
  + \la^3\, \sum_{\ell \in M} \frac{ C_{00}{}^\ell C_{\ell 0}{}^0 + C_{00}{}^\ell C_{0\ell}{}^0 }{  2y_0 (2y_0 - y_\ell)} + \ldots \; ,
  \\
  \hat{\la}_k = - \la^2 \frac{C_{00}{}^k}{2y_0-y_k}
  + \la^3\, \sum_{\ell \in M} \frac{ C_{00}{}^\ell C_{\ell 0}{}^k + C_{00}{}^\ell C_{0 \ell}{}^k }{ (2y_0 - y_\ell)(3y_0 - y_k )} + \ldots \; ,
\end{gather}
however, the combinatorics required to find the exact answer to all orders would take many pages of exposition and is unnecessary in light of a short-cut using renormalisation group technology.

\subsection{RG calculation}
\label{sec:RGintro}

Note the answer \eqref{eq:ans1} is already hinting at an RG
connection, one interprets the $\hat{\la}_k$ as
renormalised couplings replacing the bare coupling $\la$.  A second
clue is that at each order the
leading $\delta$ contributions came from the region of integration
where all the perturbing fields come together simultaneously.  In a regulated correlator, this region is cut out and it's contribution absorbed into the renormalised couplings.

Consider a more general perturbation with an $\ep$-dependent regulator,
\begin{align}
  S_\text{reg} = \sum_{k\in F} \mu_k(\ep) \ep^{-y_k} \int_{\partial \Sigma}
   d\tau \phi_k(\tau)\,,
\end{align}
which leads to the following perturbative expansion for generic
correlation functions $\langle \ldots \rangle_\mu$,
\begin{align}
  \langle \ldots \rangle_\mu
  = \langle \ldots & e^{-S_\text{reg}} \rangle_{B_\de,\ep}
  =  \langle \ldots \rangle^{}_{B_{\de}} - \sum_{k \in  F}  \mu_k(\ep) \ep^{-y_k} \int_{\partial \Sigma}
   d\tau  \langle \ldots \phi_k(\tau) \rangle_{B_{\de}}  \label{eq:regdefn}\\
  &+ \sum_{k,\ell \in F} \frac{\mu_k(\ep)\mu_\ell(\ep) \ep^{-y_k-y_\ell}}{2} \int_{\partial \Sigma}
 d\tau_1d\tau_2 \rho_\ep(\tau_1,\tau_2) \langle \ldots \phi_k(\tau_1) \phi_\ell(\tau_2) \rangle_{B_{\de}} + \ldots \nn
\end{align}
For the moment, the perturbation contains all boundary fields. ${\rho}$ denotes the regulator, we will use the usual step function $\rho_\ep(x_1,x_2) = \theta( |x_1{-}x_2|{-}\ep )$, $\rho_\ep(x_1,x_2,x_3) = \theta( |x_1{-}x_2|{-}\ep ) \theta( |x_2{-}x_3|{-}\ep ) \theta( |x_3{-}x_1|{-}\ep )$, $\ldots$.  Later we will show the result is independent of the choice of regulator.
The coupling constants in the regulated correlators are defined such
that the resulting correlator is independent of the (small) cut-off $\ep$.  To
second order in perturbation theory and using the step-function regulation we find the following renormalisation group equations,
\begin{equation} \label{eq:RG}
  \ep \frac{d \mu_k(\ep)}{d\ep} = y_k \mu_k(\ep) - \sum_{i,j \in F} C_{ij}{}^k \mu_i(\ep)\mu_j(\ep) \; .
\end{equation}
As $\ep \to 0$, the regulated correlator should reproduce the unregulated result,
\begin{equation}\label{eq:RGcond}
  \lim_{\ep \to 0} \mu_0(\ep) \ep^{-y_0} = \lambda \; ,
\qquad
  \lim_{\ep \to 0} \mu_k(\ep) \ep^{-y_k} = 0 \; ,\; k \ne 0 \; .
\end{equation}
Together these equations fix the renormalised couplings and ensure that
\begin{gather}
 \langle \ldots \rangle_{\mu, \ep} =
\langle \ldots \rangle_\la \; . \label{eq:equal}
\end{gather}
To see what this has to do with the leading $\de$ behaviour let us observe
that equation \rf{eq:equal} implies in particular that
\begin{align}
    \sum_{k \in F} \mu_k(\ep) & \ep^{-y_k} \int d\tau \langle \ldots \phi_k(\tau) \rangle  \\
   &= \la  \int d\tau \langle \ldots \phi_0(\tau) \rangle - \frac{\la^2}{2} \int d\tau_1d\tau_2 \left( 1 {-} \rho_\ep ( \tau_1,\tau_2) \right) \langle \ldots \phi_0(\tau_1) \phi_0(\tau_2) \rangle + \ldots \; ,\nn
\end{align}
The renormalised couplings contain the contribution
to the perturbed correlators cut-out by the cut-off.  Using the OPE to write all the correlators in the same form we can equate coefficients,
\begin{align}
  \mu_0(\ep) \ep^{-y_0} &= \la
  - \frac{\la^2}{2} \int du \left( 1 {-} \rho_\ep ( \tau,\tau+u) \right) C_{00}{}^0 \left( 2 \sin \tfrac u2 \right)^{-1+y_0} + \ldots  \\
  &= \la - \la^2 \frac{ C_{00}{}^0 \ep^{y_0}}{y_0} + \ldots \; , \label{eq:mu0} \\
    \mu_k(\ep) \ep^{-y_k}
    &= -\frac{\la^2}{2} \int du \left( 1 {-} \rho_\ep ( \tau,\tau+u) \right) C_{00}{}^k \left( 2 \sin \tfrac u2 \right)^{-1+2y_0-y_k} + \ldots  \\
 &= -\la^2 \frac{C_{00}{}^k \ep^{2 y_0 - y_k}}{2 y_0 - y_k} + \ldots \; . \label{eq:muk}
\end{align}
Since the leading behaviour
comes from the removed area and the removed area is encoded in the renormalised couplings, all the leading behaviour must be encoded
in the renormalised couplings.  We can extract this by taking the leading behaviour of the renormalised couplings,
\begin{align}
  \mu_0(\ep) \ep^{-y_0} &= \la - \la^2 \frac{ C_{00}{}^0}{y_0} + \ldots \; , \notag \\
  \mu_k(\ep) \ep^{-y_k} &= -\la^2\frac{C_{00}{}^k}{2 y_0 - y_k} + \ldots \; . \label{eq:rencoup}
\end{align}
which we substitute back into \eqref{eq:regdefn}.  Noting that the renormalised couplings are $\CO(\de)$ and that the integrated regulated correlators are $\CO(1)$ we reproduce the result of the brute force calculation,
\begin{align}
  \langle \ldots \rangle_\la = \langle \ldots \rangle_{\mu,\ep}
  =  \langle \ldots \rangle^{}_{B_{\de}} - \sum_{k \in  F}  \mu_k(\ep) \ep^{-y_k} \int_{\partial \Sigma}
   d\tau  \langle \ldots \phi_k(\tau) \rangle_{A} + \text{subleading in $\delta$} \; ,
  \label{eq:finalren}
\end{align}
wherein the renormalised couplings are given by \eqref{eq:rencoup}


We observe that there is a simplification that we can make.
Since we are  only interested in the leading $\delta\ra 0$
behaviour of the
renormalised couplings, looking at equations \eqref{eq:RG}
and solving them as power series in
$\lambda$ one can see this is encoded in
the subset of equations involving only the almost marginal couplings,
with all other couplings set to zero.
This is the renormalisation group realisation of the fact that when we
did the brute force calculation, we needed
only the part of the OPEs involving the almost marginal fields.

Let us summarise our arguments.  As seen from the brute force
calculation, the leading $\delta$ behaviour of correlators came from
the regions of integration where the all perturbing fields came
together simultaneously.
In a suitable renormalisation scheme, the contribution from these regions is cut-out and absorbed into the renormalised couplings.
It should be emphasized that the expansion in powers of the renormalized
coupling constants $\mu$ is simply a reorganization of the
perturbative expansion in powers of $\la$, as is clear from
equation \rf{eq:equal}.
One can efficiently calculate these couplings by
solving the RG equations with correct boundary conditions.
The power of the method is illustrated by
the fact that equations \eqref{eq:mu0} and
\eqref{eq:muk} can also be easily calculated from the
renormalisation group equations \eqref{eq:RG}.
The
leading $\delta$ behaviour of the solutions to the
RG equations will then give the
leading $\delta$ behaviour of the correlators via \eqref{eq:finalren}.

\subsection{Toy Example : a single nearly marginal field}
\label{sec:toy}

As an illustration of our method, we consider the
perturbation in a theory with a single nearly marginal field, $y=1{-}h=\delta^2$.  We use $\de^2$ to make comparison with the time-dependent formulation is easier.
\begin{align} \label{eq:toypert}
  \langle \ldots \rangle_\la = \langle \ldots e^{-S_\text{pert}} \rangle^{}_{B_{\de}} \; ,
  \qquad
  S_{\text{pert}} = \la \int d\tau \phi(\tau) \; .
\end{align}
In this case the RG equations are,
\begin{align}
  \ep \frac{d \mu(\ep)}{d\ep} = \delta^2 \mu(\ep) - C  \mu(\ep)^2 \; ,
  \qquad
  \lim_{\ep \to 0} \mu(\ep) \ep^{-\delta^2} = \lambda \; ,
\end{align}
which solve to give,
\begin{align}
  \mu(\ep) \ep^{-\delta^2} = \frac{\delta^2 \la }{\delta^2 + C \ep^{\delta^2} \la} = \frac{\delta^2 \la }{\delta^2 + C \la} + \text{subleading in $\delta^2$} \; ,
\end{align}
and hence the leading $\delta^2$ behaviour of the perturbed correlator is,
\begin{align}\label{pertcorr:ex}
  \langle \ldots \rangle_\la =  \langle \ldots \rangle^{}_{B_{\de}} - \frac{\delta^2 \la }{\delta^2 + C \la}
  \int d\tau \langle \ldots \phi(\tau) \rangle^{}_{B_{\de}}
  + \text{subleading in $\delta$} \; .
\end{align}
In the present example it is possible to
carry out the  brute force calculation
by expanding \eqref{eq:toypert} in powers of $\la$,
using the OPE to calculate the leading $\delta$
behaviour term by term and then resumming the series. The result
coincides with \rf{pertcorr:ex}.

Note that the series \eqref{eq:toypert} has a finite radius of convergence.
By performing the resummation we obtain a
continuation to all values of $\la>0$ (for $\la<0$, the renormalised coupling becomes large and
our perturbative assumptions break down). This is particularly
important when one tries to calculate the
perturbed correlation function at a nearby renormalization group fixed point.
In this case the renormalized coupling constant stays small and
approaches the fixed point value
$\mu(\ep) \to \tfrac{\de^2}{C}$
as $\la \to \infty$. The renormalized coupling constant captures
information that is nonperturbative in $\la$.

\subsection{Scheme dependence}
\label{sec:schemeind}

The equations above have been derived using a step-function cut-off.  In this section we consider a more general cut off $\rho_\ep(x_1,x_2) = \rho( \tfrac 1\ep |x_1-x_2| ) $.  Re-deriving the RG equations we find,
\begin{equation}
   \ep \frac{d \mu_k(\ep)}{d\ep} = y_k \mu_k(\ep) - \sum_{i,j \in M}
C_{ij}{}^k \mu_i(\ep)\mu_j(\ep) f(y_i+y_j-y_k),
   \quad
  f(y) = \int_0^\infty du \, u^y \frac{d \rho(u)}{du}.
\end{equation}
What is important for our calculation is the leading $\delta$ behaviour, and one will note that since $\rho$ is a cut-off function (whose derivative is well behaved) we have
\begin{align}
  f(\delta) &= \rho(\infty)-\rho(0) + \text{subleading in $\delta$} \notag \\
  &= 1  + \text{subleading in $\delta$}\,.
\end{align}
Hence as far as the leading $\delta$ behaviour is concerned, the result
is scheme independent for all schemes whose cut-off cleanly removes all the
behaviour from the integrals which would become divergent when $\de\ra 0$.
Such schemes will subsequently be called ``proper''\footnote{
Using the equations like those  derived in this section one could easily give some sufficient conditions
for the definition of such a scheme. }.

\section{Perturbative construction of time-dependent backgrounds}
\label{TsummingRG}

The string backgrounds that we are interested in can be constructed
as perturbations of the product
conformal field theory ${\rm CFT}_{\rm S}\ot{\rm CFT}_{\rm X_0}$.
where ${\rm CFT}_{\rm S}$ is a conformal field theory with central
charge $c=25$ representing the spatial part of the background
and
${\rm CFT}_{\rm X_0}$ is a
free boson CFT that is defined by the action
\begin{equation}\label{WSact}
S\,=\,-\frac{1}{4\pi}\int d^2
x \; \pa_+ X_0\pa_- X_0\, . \end{equation}
The sign in front of the action means that $X_{0}$ is
time-like.

We will be interested in
certain perturbations of the static open string backgrounds which are
characterized by a family of conformal
boundary conditions $B_{\de}^{}$ for ${\rm CFT}_{\rm S}$
parametrized by a parameter $\de$ together with
Neumann boundary conditions for the $X_0$-CFT.
The boundary state for the static background
can be written as \begin{equation} |\,B_{\de}^{}\,\kket_{\rm
stat}^{}\,\equiv\, |\,B_{\de}^{}\,\ket^{}_{\rm S}
\ot|\,N\,\xket\,, \end{equation}
where $|\,N\,\xket$ is the boundary state associated to Neumann type
boundary condition for the $X_0$-CFT.
The boundary state of the time-dependent (dynamical) background
that we are about to study will be denoted as
$|\,B_{\de}^{}\,\kket_{\rm dyn}^{}$.
It may be formally constructed as follows,
\begin{equation}\label{pertB}
|\,B_{\de}^{}\,\kket_{\rm dyn}^{}\,\equiv\, e^{-S_{\rm
Bd}}\,|\,B_{\de}^{}\,\ket^{}_{\rm S}\ot|\,N\,\xket\,, \end{equation}
where $S_{\rm Bd}$ is the
following boundary action:
\begin{equation}\label{Abare} S_{\rm Bd}\,\equiv\,\la
\int_{\pa\Sigma}dx\; [e^{\de X_0}\phi_0](x)\,.
\end{equation}
here we have fixed $y_0 = \de^2$ such that the perturbing field
has conformal dimension $\de^2+(1-\de^2)=1$.
There are cases where the short distance singularities in
the OPE of $[e^{\de X_0}\phi_0](x)$
with itself are integrable.
It then follows that \rf{pertB}
indeed defines a conformal boundary state
to all orders in a formal expansion in the
parameter $\la$.

We observe an immediate problem: The perturbative expansion
in $\la$ is not expected to be convergent since a shift of the
zero mode of $X_0$ is equivalent to a rescaling of $\la$.
In the following subsection we will describe how this problem
may be solved by fixing the zero mode value.
We will then explain how to
calculate the boundary state
$|\,B_{\de}^{}\,\kket_{\rm dyn}^{}$
to leading order in $\de$ by generalizing the RG techniques
from the previous section to the time-dependent case.

In order to explain our method we will assume that the
short distance singularities in
the OPE of $[e^{\de X_0}\phi_0](x)$
with itself are integrable. This assumption is made for
pedagogical purposes only, our main conclusions do not depend on it,
as is shown in Appendix A.

\subsection{The $X_0$ BCFT}

To begin with, let us consider the $X_0$-CFT on the cylinder with
periodic boundary conditions in  space direction $\si$.
The space of states of the $X_{0}$ theory is generated from
a continuum of states $\langle
\omega|_{X_{0}}$, $\omega\in {\mathbb R}$ satisfying
\begin{equation} \xbra{\omega}\,L_{-n}=0=\xbra{\omega}\,\bar{L}_{-n},\quad
n>0,\quad \xbra{\omega}\,L_0\,=\,-\omega^2\xbra{\omega}\, ,
\end{equation} where $L_{n}$, $\bar{L}_n$
are the generators of the $c=1$ Virasoro
algebra.

In order to characterize
the perturbed boundary state
$|\,B_{\de}^{}\,\kket_{\rm dyn}^{}$ one would naturally
consider the amplitude
\begin{equation}\label{Aawdef}
A(a,\om)\,\equiv\,
\bbra \,a,\om\, |\,B_{\de}^{}\,\kket_{\rm dyn}^{}\,,\qquad \bbra
\,a,\om\,|\,\equiv\,\langle \,a\,|_{\rm S}^{}\ot\xbra{\om}\,,
\end{equation}
where $\langle \,a\,|_{\rm S}^{}$ is a highest weight state in
the ${\rm CFT}_{\rm S}$ on the cylinder.
However, as indicated in the introduction to this section
one can not expect that the perturbative expansion in powers
of $\la$ will be useful to determine $A(a,\om)$.

In order to overcome this problem, let us
introduce the zero mode $x_0= \int_{0}^{2\pi} d\si\;
X_0(\si,0)$. We may fix the zero mode $x_0$ by considering amplitudes which
involve the eigenstates $\xbra{t}$ of $x_0$. The states $\xbra{t}$
 are obtained from $\xbra{\omega}$ by Fourier-transformation,
\begin{equation} \xbra{t}\,=\,\int_\BR d\omega\; e^{i\omega
t}\,\xbra{\omega}\,. \end{equation}
It follows that the states $\xbra{t}$ satisfy
\begin{equation}\label{teigen} \xbra{t}\,L_{-n}=0=\xbra{t}\,\bar{L}_{-n},
\quad n>0,\quad \xbra{t}\,x_0\,=\,t\,\xbra{t}\,. \end{equation}
Instead of considering $A(a,\om)$,
we will first determine the amplitude
\begin{equation}\label{Aatdef}
A(a,t)\,\equiv\,\bbra \,a,t\, |\,B_{\de}^{}\,\kket_{\rm dyn}^{}\,,\qquad \bbra
\,a,t\,|\,\equiv\,\langle a|^{}_{\rm S}\ot\xbra{t}\,.
\end{equation}
We will interpret  $A(a,t)$ as an amplitude which directly
represents the time evolution of the boundary state
$|\,B_{\de}^{}\,\kket_{\rm dyn}^{}$.

The perturbative expansion for $A(a,t)$ in powers of $\la$ involves
amplitudes like
\begin{equation}
\bra \,t\,|:\!e^{\de X_0(\si_1)}\!:\ldots :\!e^{\de X_0(\si_n)}\!:|\,N\,\rangle_{X_{0}}^{}
\;=\;e^{n\de t}\,\prod_{r<s}\left|\,2\sin\frac{\si_r-\si_s}{2}\,\right|^{2\de^2}\,.
\end{equation}
The prefactor $e^{n\de t}$ comes from the
zero mode dependence of the normal ordered exponentials
$e^{\de X_0(\si)}$ together with \rf{teigen}. The $\si_r$-dependent
factors follow from the OPE
\[
:\!e^{\de X_0(\si_2)}\!::\!e^{\de X_0(\si_1)}\!:\,=\,
\Big|\,2\sin\frac{\si_2-\si_1}{2}\,\Big|^{2\de^2}:\!
e^{\de X_0(\si_2)}e^{\de X_0(\si_1)}\!:\,,
\]
which is the usual OPE of normal ordered exponentials of a free
field up to the change of sign in the exponent of the short-distance
factor due to the minus sign in front of the kinetic term for
$X_0$.
The time-like nature of the $X_0$-CFT is directly responsible for the
fact that the short-distance behavior of the
operator product $e^{\de X_0(\si_2)}e^{\de X_0(\si_1)}$ is {\it nonsingular}.
It follows that
\begin{equation}
\bra \,t\,|:\!e^{\de X_0(\si_1)}\!:\ldots :\!e^{\de X_0(\si_n)}\!:|\,N\,\rangle_{X_{0}}^{}
\;=\;e^{n\de t}\,(1+\CO(\de^2))
\end{equation}
holds as long as $n$ is not of the order $\de^{-1}$.

With these definitions one may use equations \rf{pertB} and \rf{Aatdef} to
generate a formal series expansion
of $A(a,t)$ in powers of $\lambda e^{\de t}$.
We expect that the radius of convergence of this expansion is
finite, as will be confirmed below.
We will nevertheless be able to find explicit representations for
$|\,B_{\de}^{}\,\kket_{\rm dyn}^{}$ and $A(a,t)$
which are valid to leading order in $\de$ but for
arbitrary  $\la e^{\de t}$ by
using the renormalization group resummation of the
naive perturbative expansion in
powers of $\la$ as discussed in Section \ref{summingRG}.

\subsection{RG improvement in the time-dependent case}\label{TsummingRG2}

To represent the decay process in string theory, we couple our CFT to
a time-like boson with a Neumann-type boundary condition.  We then
perturb the theory by the (truly) marginal operator,
\begin{equation}
  S= \lambda \int dx \;[ e^{\delta X^0} \phi_0 ] (x) \; .
\end{equation}
with $y_0 = \delta^2$.
Using the technology above, this leads us to consider the regulated
perturbation,
\begin{equation}
  S_{\text{reg}} =\sum_{k \in M} \sum_{n=1}^\infty u_{k,n}
\ep^{n^2\delta^2-y_k} \int dx\;
  [ e^{n \delta X^0} \phi_k ] (x)\,. \label{eq:regAction}
\end{equation}
One may wonder why we did not expand $U(X_{0})$ and $\la_{P}(X_{0})$ into
``Fourier modes'' $:\!e^{-i\om X_0}\!:$, as has been done
e.g. in \cite{FHL}. In this case one would find fields in the
boundary action with arbitrarily negative conformal dimensions, which would
in particular create problems in the application of RG techniques.
We have included all the near marginal boundary fields which are
generated in the
OPE of the perturbing field
$[e^{\de X_0}\phi_0](x)$ with itself.

The conditions for the $\ep$-independence of the
correlation functions are then found to be
\begin{gather}\label{eq:tRG}
  \ep \frac{d u_{k,n}}{d\ep} =  (y_k- n^2 \delta^2) u_{k,n} -\sum_{i,j\in M}
  \sum_{m=1}^{n-1} C_{ij}{}^k u_{i,m} u_{j,n-m} \; , \\
  \lim_{\ep \to 0}  u_{0,1} = \la \; ,
  \qquad
  \lim_{\ep \to 0}  u_{k,n} \ep^{n^2\delta^2-y_k} = 0 \; .
\end{gather}
These equations can be solved recursively,
\begin{equation}
   u_{0,1} = \la \; ,
  \qquad
   u_{k,1} = 0 \; ,
  \qquad
   u_{k,n} = \frac 1{y_k- n^2 \delta^2} \sum_{i,j\in M}
  \sum_{m=1}^{n-1} C_{ij}{}^k u_{i,m} u_{j,n-m}\,.
\end{equation}
This  is also a fixed point of \eqref{eq:tRG}, as was to be
expected because our original perturbation was truly marginal.
The equations above can be translated into a system
of evolution equations by introducing
\begin{align}\label{Ukdef}
  U_k(t) = \sum_{n=1}^\infty u_{k,n} e^{n \delta t}\,.
\end{align}
It easily follows that $U_k(t)$ is a solution of the equations
\begin{align}\label{Uevol}
  \ddot{U}_k(t) = y_k U_k(t) - \sum_{i,j \in M} C_{ij}{}^k U_i(t)
  U_j(t) \,
\end{align}
supplemented by the boundary conditions
\begin{equation}\label{Ubc}
U_0(t)\,=\,\la e^{\de t}+\CO(e^{2\de t})\,,\qquad
U_k(t)\,=\,\CO(e^{2\de t})\,,\quad k \ne 0\,.
\end{equation}
When we compare this system with the renormalization group
equations \rf{eq:RG} of the time-independent treatment, we clearly
would not expect any simple relation between the solutions to the
respective equations.
We will see an explicit example in a moment.

The leading behaviour for $\de\ra 0$ of time dependent correlators
$\langle \ldots \rangle_\la$
is then given as
\begin{align}
  \big\langle \ldots \big\rangle_\la \,=\,
\big\langle \ldots
\big\rangle_{B_{\de} \ot N}^{} -
\sum_{k \in F} \sum_{n=1}^\infty u_{k,n} \ep^{n^2 \de^2-y_k}
\int d\tau \;\big\langle
\ldots [e^{n \delta X_0} \phi_k](\tau)
\big\rangle_{B_{\de} \ot N}^{} \; .
\end{align}
Of particular interest is the amplitude $A(a,t)$ defined
in  equation \rf{Aatdef} above. The leading behavior of this quantity
may be represented as
\begin{align}
A(a,t) &\dasym
\bbra \,a,t\, |\,B_{\de}^{}\,\kket_{\rm stat}^{}
- \sum_{k \in F} \sum_{n=1}^\infty u_{k,n} \ep^{n^2 \de^2-y_k}
\int d\tau\;
\bbra \,a,t\, |\,
[e^{n \delta X_0}\phi_k](\tau) \,|\,B_{\de}^{}\,\kket_{\rm stat}^{}
\nn  \\
  &\dasym \bra \,a\,|\,B_{\de}\,\rangle_{\rm S}^{} -
\sum_{k \in F} U_k(t) \int d\tau \;\bra \,a\,|\, \phi_k(\tau)\,
|\,B_{\de}\,\rangle_{\rm S}^{}\,.
\label{Aatrepr}\end{align}
We observe that to leading order in $\de$ one may represent
the time-dependence of the perturbed amplitudes $A(a,t)$ rather
simply in terms of the solutions to the evolution equations \rf{Uevol}.

A few comments are at order at this point. A priori we may expect
the representation \rf{Aatrepr} to be useful only for times $t$
which are sufficiently small to ensure the convergence of the series
\rf{Ukdef}. The possibility to find representations for the amplitude
$A(a,t)$ valid for all times $t$ depends crucially on whether the
dynamics defined by the equations \rf{Uevol},\rf{Ubc} will remain
bounded or not. If not, one would violate the condition that $U_k=\CO(\de)$
after some time. The possibility of unbounded motions is raised
by the fact that the right hand side of \rf{Uevol} is the force from
a cubic potential which is unbounded from below. We will not be
able to offer a general answer to this question, but we will
prove boundedness of the dynamics defined by \rf{Uevol},\rf{Ubc}
in two interesting cases below.

If the functions $U_k(t)$ have a well defined limit as $t\to
\infty$ it necessarily has to be a fixed point of the time-independent RG
flow.   However, without more information, there is no reason to
expect this fixed point will be the same as the the end
point of the RG flow that is generated by the boundary field $\phi_0$.
We will indeed illustrate in the next subsection that this is not
the case in general.

Let us finally remark that
both the evolution equations and the RG equation for the time independent
system are scheme dependent in general.  Furthermore, there is no reason to
suppose that a scheme chosen for one system should be related to a
scheme chosen for the other in a simple way.
However, as argued in section
\ref{sec:schemeind},
the {\it leading behaviour} of both systems for $\de\ra 0$ is
{\it scheme
independent}.

\subsection{Toy Example : a single marginal field}
\label{sec:timetoy}

We continue our example from section \ref{sec:toy}.  To
create the time-dependent version of this system, we couple our single near
marginal field to the time-like boson,
\begin{equation}
  S= \lambda \int dx \;[ e^{\delta X^0} \phi] (x) \; .
\end{equation}
Using the technology of RG, this leads us to consider the more general
perturbation,
\begin{equation}
  S_{\text{reg}} = \sum_{n=1}^\infty u_{n} \ep^{(n^2-1)\delta^2} \int dx
 [ e^{n \delta X^0} \phi ] (x) \; ,
\end{equation}
and the RG equations,
\begin{gather}\label{eq:timeRG}
  \ep \frac{d u_{n}}{d\ep} =  (1- n^2) \delta^2 u_{n} -
  \sum_{m=1}^{n-1} C u_{m} u_{n-m} \; ,
  \qquad
  \lim_{\ep \to 0}  u_{1} = \la \; ,
  \qquad
  \lim_{\ep \to 0}  u_{n} \ep^{n^2\delta^2-y_k} = 0 \; .
\end{gather}
Which have the solution
\begin{equation}
   u_n = (-1)^{n-1} n\lambda \left(\frac{ C\lambda }{6\delta^2} \right)^{n-1}\, .
\end{equation}
To understand the solution better we consider the combination
\begin{align}
  U(t) = \sum_{n=1}^\infty u_{n} e^{n \delta t}
  = \frac{3y}{2C} \frac{1}{\cosh^2{\delta(t{-}t_0)}} \; , \qquad e^{\delta
    t_0} = \frac{6y\lambda}{C}\, .
\end{align}
Which satisfies
\begin{align}
  \ddot{U}(t) = \delta^2 U(t) - C\, U(t)^2 \; .
\end{align}
This is the equation of motion for a particle in a cubic potential.
The solution above represents the particle leaving $U=0$ in the
infinite past, falling toward the minimum $U_* = \tfrac{\delta^2}{C}$,
moving on up the other side before coming to instantaneous rest at a
point $U(t_0)=\tfrac 32 U_*$. The particle then returns to $U=0$ in the
infinite future.

For systems with only a finite number of nearly marginal fields, such
oscillatory behaviour will be generic.  To find a system of a
dissipative nature, we need an infinite number of fields, as is found
in the example that we will consider next.

\section{(Perturbed) Liouville BCFT} \label{LBCFT}
\setcounter{equation}{0}

Liouville theory on the upper half plane $\BU$
is defined semiclassically
by means of the action
\begin{equation}
S_{\rm L}[\phi]\,\equiv\,\frac{1}{\pi}\int_{\BU}d^2z\left(\pa\phi
\bar{\pa}\phi+\pi\mu e^{2b\phi}\right)
+\mu_B\int_{\BR}dx\;e^{b\phi}\,.
\end{equation}
The corresponding boundary conditions for the Liouville field
are of Neumann-type,
\begin{equation}\label{classBC}
i(\pa-\bar\pa)\phi\,=\,2\pi b\, \mub\, e^{b\phi}\,.
\end{equation}
One of the interesting implications of the exact solution
of boundary Liouville theory \cite{FZZ,T1,Ho,PT,T2} is the fact that the
boundary conditions of the corresponding quantum theory
are not uniquely parametrized by the
parameter $\mub$ which appears in the classical case \rf{classBC}.
For each value of $\mub$ there are countably many different boundary
conditions which have quite different physical properties. In the
following we will elaborate on the string theoretical
consequences of this phenomenon, extending the previous discussion in
\cite{T2}.

We will exclusively consider the case $b=1$ in this paper,
which corresponds to central charge $c_{\rm L}=25$.
The primary fields of the theory are distinguished by their conformal weights, $\Delta$,
and will be denoted by $V_{\al}(\tau,\si)$ in the bulk and $\Phi_P(x)$ on the boundary.
The labels $P$ and $\al$ will be used interchangably and are related to each other and the conformal weight via
$\Delta_{\alpha} = 1+P^2 = \al(2-\al)$, $\al = 1+iP$.



\subsubsection{Boundary state}

The boundary states which correspond to the classical boundary
condition \rf{classBC} were first presented in \cite{FZZ}.
They can be represented as 
Ishibashi-states $|P\ket^{}_{\rm Ish}$,
\begin{equation}\label{boundst}
|B_\de^{}\ket^{}_{\rm L}
\;=\;\int_0^{\infty}\frac{dP}{2\pi}\;A^{{\de}}_{P}\;
|P\ket^{}_{\rm Ish}\;
\end{equation}
where $|P\ket^{}_{\rm Ish}$ is the Ishibashi state built upon the bulk Liouville heighest weight
state $|P\rangle$ corresponding to the vertex operator $V_{\alpha}$.
The one-point function $A^{{\de}}_{P}$ depends on a parameter $\de$
which is related to the
boundary cosmological constant $\mub$ via\footnote{Other parametrizations
have been used in the literature. $\de$ is related to the parameter
$\si$ from \cite{PT,T2} as $2\si=1-\de$, whereas the parameter
$s$ from \cite{FZZ} is related to $\de$ as $s=i(1+\de)$.}
\begin{equation}\label{smub}
\cos \pi (1+\de)
=\frac{\mub}{\sqrt{\mu}}\;.
\end{equation}
The explicit expression for the coefficients $A^{{\de}}_{P}$ is then
given as
\begin{equation}\label{onept}
A^{{\de}}_{P}\;=\;
\frac{\cosh(2\pi P(1+\de))}{2\,(\sinh 2\pi P)^2}
\,\Theta(P),\qquad
\Theta(P)\,:=\, 
\frac{4\pi iP\,
\mu^{{iP}}}{(\Ga(1+2iP))^2}\,.
\end{equation}
It has turned out to be useful to split off
the function  $\Theta(P)$ as a normalizing factor.

We will mainly be interested in the case of small values of $\de$,
which corresponds
to the region around the first minimum of $\mub=\mub(\de)$ on the real
positive half-axis. Bear in mind that
$-\de$ corresponds to the same value of $\mub$.


\subsection{Hamiltonian picture - closed string channel}\label{LHampic}

We would like to understand the qualitative differences between
the cases $\de>0$ and $\de<0$.
Some useful insight can be obtained by considering the Hamiltonian picture
for Liouville theory which is naturally associated to the
world-sheet being the cylinder.
The boundary state is considered as description for the initial state at
$\tau=0$. We may consider expectation values like
\begin{equation}
\bra \,0\,|\,V_{\al_n}(\tau_n,\si_n)\dots
V_{\al_\1}(\tau_\1,\si_\1)\,|\,B_{\de}^{}\,\lket\,.
\end{equation}

It is natural to interpret
the zero mode $\phi_0\equiv\int_0^{2\pi}d\si \phi(\si)$
as a coordinate for the target space of Liouville theory.
In order to discuss localization properties in the
target space
it is useful to think about the states in $\CH_{\rm L}$
in terms
of the Schr\"odinger representation\footnote{The zero mode
$\phi_0$ is an operator which can be
constructed from the exponential fields of
Liouville theory \cite{TL}. It is unbounded,
but well-defined on a dense domain,
symmetric,  $\phi_0^\dagger=\phi_0$,
and it seems likely that $\phi_0$ has a self-adjoint extension.}
for the zero mode
$\phi_0$. States $|\Psi\lket$ are then
represented by wave-functions $\Psi(\phi_0)\in\CH(\phi_0)$, and the norm
$\lVert \Psi\rVert^2$ is represented in the form
\begin{equation}
\lVert\, \Psi\,\rVert^2\,\equiv\,\int d{\phi_0}\;\lVert\,\Psi({\phi_0})\,
\rVert^{2}_{\CH(\phi_0)\,.}
\end{equation}
The norm density $\lVert\,B_\de({\phi_0})\,
\rVert^{2}_{\CH(\phi_0)}$ of the wave-function associated to the
boundary state $|\,B_{\de}^{}\,\lket$ can then be seen as describing
the ``profile'' of the D-branes associated to the boundary
condition with label $\de$. At present we do not know how to
calculate these profiles explicitly, but the asymptotic behavior
can be read off from the asymptotic behavior of the one-point function
for $P\ra 0$ and $P\ra \infty$ respectively.
In order to see this let us consider
the representation of $A^\de_P$ as an overlap,
\begin{equation}\label{bdoverlap}
A^\de_P\,=\,\bra\, P\, |\,B_{\de}^{}\,\lket\,=\,
\int d{\phi_0}\;\bra\, \Psi_P({\phi_0})\,|\,B_{\de}^{}({\phi_0})\,
\ket_{\CH(\phi_0)}^{}\,.
\end{equation}
Let us first consider the asymptotics $\phi_0\ra -\infty$.
One should keep in mind  that
the wave-function $\Psi_P({\phi_0})$ behaves as \cite{ZZ,TL}
\begin{equation}
\Psi_P({\phi_0})\,\underset{{\phi_0}\ra \infty}{\sim}\,
\left(\,e^{2iP{\phi_0}}+R(P)e^{-2iP{\phi_0}}\,\right)\Om\,,
\end{equation}
where $\Om$ is the Fock vacuum, and $R(P)$ is the reflection amplitude.
The divergence found in
the wave-function of the boundary state when $P\ra 0$,
\begin{equation}\label{A0asym}
|\,A^{{\de}}_{P}\,|\,
\underset{P\ra 0}\sim\,
\frac{1}{2\pi P}
\end{equation}
is most naturally explained if
$B_\de({\phi_0})$ approaches a constant for ${\phi_0}\ra-\infty$.

In order to discuss the
asymptotics of $B_{\de}^{}({\phi_0})$ for $\phi_0\ra +\infty$
let observe that these asymptotics are
related to the asymptotics for $P\ra\infty$
of $A^\de_P$. Indeed, for large $P$ one may expect
that the rapid oscillations of the wave function $\Psi_P({\phi_0})$
will average out the contributions to the integral \rf{bdoverlap} from
a large range of values of $\phi_0$. This range is roughly bounded
from above by the turning point of
the motion of a string in the purely
repulsive Liouville potential. The purely repulsive nature of the potential
furthermore implies that $\Psi_P({\phi_0})$ will decay rapidly for
$\phi_0\ra +\infty$. It follows that
the main contributions to the integral
\rf{bdoverlap} come from the region around the
turning point of
the motion of a string in the Liouville potential.
The latter will grow with $P$.
The asymptotics of $|A^\de_P|$ for large $P$
\begin{equation}\label{inftyas}
|\,A^{{\de}}_{P}\,|\,=\,
\frac{\sinh(2\pi(1+\de)P)}{\sinh(2\pi P)}\,\underset{P\ra\infty}\sim\,
e^{2\pi \de P}\,,
\end{equation}
therefore reflects
the asymptotics of the wave-function
$B_{\de}^{}({\phi_0})$ for ${\phi_0}\ra\infty$.
Note in particular that
the latter depends decisively on the sign of $\de$.

The resulting picture looks as follows:
For each value of $\mub$ we find two different boundary states
with $-1<\de<1$
distinguished by the value of $\sgn(\de)$,
one of which ($\sgn(\de)>0$) has a strong growth of the
profile $M({\phi_0})\equiv\lVert B_{\de}^{}({\phi_0})\rVert^2$
for ${\phi_0}\ra\infty$.
In string theoretic terms one may interpret this fact as
the existence of a concentration of ``mass'' (in the sense of
source for closed strings) for large values of ${\phi_0}$ on
those D1 branes which have $\de>0$.
An intuitive way to visualize the profiles for the two cases
$\de<0$ and $\de>0$ is given in figures \ref{profile1} and
\ref{profile2} respectively.
\begin{figure}[htb]
\centering
\setlength{\unitlength}{0.00047489in}
\begingroup\makeatletter\ifx\SetFigFont\undefined%
\gdef\SetFigFont#1#2#3#4#5{%
  \reset@font\fontsize{#1}{#2pt}%
  \fontfamily{#3}\fontseries{#4}\fontshape{#5}%
  \selectfont}%
\fi\endgroup%
{\renewcommand{\dashlinestretch}{30}
\begin{picture}(7674,2199)(0,-10)
\path(12,372)(7662,372)
\blacken\path(7542.000,342.000)(7662.000,372.000)(7542.000,402.000)(7542.000,342.000)
\path(4062,372)(4062,2172)
\blacken\path(4092.000,2052.000)(4062.000,2172.000)(4032.000,2052.000)(4092.000,2052.000)
\path(4062,372)(4062,12)
\path(12,1452)(13,1452)(17,1452)
	(23,1452)(33,1452)(47,1452)
	(66,1452)(90,1452)(121,1452)
	(157,1451)(198,1451)(246,1451)
	(299,1451)(358,1450)(421,1450)
	(489,1450)(561,1449)(635,1449)
	(712,1448)(791,1448)(871,1447)
	(952,1446)(1033,1446)(1113,1445)
	(1193,1444)(1271,1443)(1348,1443)
	(1423,1442)(1497,1441)(1568,1440)
	(1637,1439)(1704,1438)(1768,1437)
	(1830,1435)(1890,1434)(1948,1433)
	(2004,1431)(2058,1430)(2110,1429)
	(2161,1427)(2210,1425)(2257,1424)
	(2303,1422)(2348,1420)(2392,1418)
	(2435,1416)(2477,1414)(2519,1412)
	(2559,1409)(2599,1407)(2600,1407)
	(2647,1404)(2694,1401)(2740,1398)
	(2786,1394)(2832,1390)(2878,1387)
	(2924,1383)(2970,1378)(3016,1374)
	(3061,1369)(3107,1365)(3153,1360)
	(3199,1354)(3244,1349)(3290,1344)
	(3335,1338)(3381,1332)(3426,1326)
	(3471,1320)(3515,1313)(3559,1307)
	(3603,1300)(3647,1293)(3690,1286)
	(3732,1279)(3774,1272)(3815,1265)
	(3856,1258)(3896,1250)(3935,1243)
	(3974,1236)(4012,1228)(4049,1221)
	(4086,1213)(4122,1206)(4157,1198)
	(4192,1190)(4227,1183)(4261,1175)
	(4295,1168)(4329,1160)(4362,1152)
	(4399,1143)(4436,1134)(4473,1125)
	(4510,1116)(4548,1107)(4586,1097)
	(4624,1087)(4663,1077)(4702,1067)
	(4742,1057)(4782,1046)(4823,1036)
	(4863,1025)(4905,1014)(4946,1003)
	(4987,992)(5029,981)(5071,969)
	(5112,958)(5154,947)(5195,936)
	(5236,924)(5277,913)(5317,902)
	(5357,892)(5396,881)(5435,871)
	(5473,860)(5511,850)(5548,841)
	(5584,831)(5620,822)(5655,812)
	(5690,803)(5724,795)(5758,786)
	(5791,778)(5825,769)(5861,760)
	(5898,752)(5935,743)(5972,734)
	(6010,726)(6047,717)(6085,709)
	(6123,700)(6161,692)(6200,684)
	(6238,675)(6277,667)(6316,659)
	(6355,652)(6393,644)(6431,636)
	(6469,629)(6507,622)(6543,615)
	(6580,608)(6615,601)(6650,595)
	(6684,589)(6717,583)(6749,577)
	(6780,572)(6810,567)(6839,562)
	(6867,557)(6894,553)(6920,549)
	(6946,545)(6971,541)(6995,537)
	(7027,532)(7059,527)(7090,523)
	(7121,518)(7151,514)(7182,510)
	(7212,506)(7244,502)(7277,499)
	(7310,495)(7344,491)(7379,487)
	(7415,483)(7449,479)(7483,476)
	(7515,472)(7543,469)(7567,467)
	(7586,465)(7600,464)(7610,463)
	(7615,462)(7617,462)
\put(4242,1902){\makebox(0,0)[lb]{\smash{{{\SetFigFont{12}{34.8}{\rmdefault}{\mddefault}{\updefault}$M$}}}}}
\put(7537,62){\makebox(0,0)[lb]{\smash{{{\SetFigFont{12}{34.8}{\rmdefault}{\mddefault}{\updefault}$\phi_0$}}}}}
\end{picture}
}
\caption{Qualitative visualization of the D1 brane profile $M({\phi_0})$
for $\de<0$. }\label{profile1}
\end{figure}
\begin{figure}[htb]
\centering
\setlength{\unitlength}{0.00047489in}
\begingroup\makeatletter\ifx\SetFigFont\undefined%
\gdef\SetFigFont#1#2#3#4#5{%
  \reset@font\fontsize{#1}{#2pt}%
  \fontfamily{#3}\fontseries{#4}\fontshape{#5}%
  \selectfont}%
\fi\endgroup%
{\renewcommand{\dashlinestretch}{30}
\begin{picture}(7674,4089)(0,-10)
\path(12,462)(7662,462)
\blacken\path(7542.000,432.000)(7662.000,462.000)(7542.000,492.000)(7542.000,432.000)
\path(4062,462)(4062,4062)
\blacken\path(4092.000,3942.000)(4062.000,4062.000)(4032.000,3942.000)(4092.000,3942.000)
\path(4062,462)(4062,12)
\path(12,1542)(13,1542)(16,1542)
	(21,1542)(29,1542)(41,1542)
	(58,1542)(78,1542)(105,1542)
	(136,1542)(173,1542)(216,1542)
	(264,1543)(318,1543)(377,1543)
	(441,1543)(510,1543)(584,1543)
	(661,1544)(742,1544)(825,1544)
	(911,1544)(999,1545)(1088,1545)
	(1178,1546)(1269,1546)(1359,1546)
	(1449,1547)(1538,1547)(1626,1548)
	(1713,1548)(1798,1549)(1882,1549)
	(1964,1550)(2044,1550)(2122,1551)
	(2198,1552)(2272,1552)(2343,1553)
	(2413,1554)(2481,1554)(2547,1555)
	(2612,1556)(2674,1557)(2735,1557)
	(2795,1558)(2853,1559)(2910,1560)
	(2966,1561)(3020,1562)(3074,1563)
	(3127,1564)(3179,1565)(3230,1567)
	(3281,1568)(3332,1569)(3382,1571)
	(3432,1572)(3490,1574)(3547,1575)
	(3604,1577)(3662,1579)(3719,1581)
	(3776,1583)(3834,1585)(3891,1588)
	(3949,1590)(4007,1592)(4065,1595)
	(4123,1598)(4182,1600)(4240,1603)
	(4299,1606)(4357,1609)(4416,1613)
	(4474,1616)(4532,1619)(4590,1623)
	(4648,1627)(4706,1630)(4763,1634)
	(4820,1638)(4876,1642)(4932,1647)
	(4987,1651)(5041,1655)(5095,1660)
	(5147,1664)(5199,1669)(5250,1673)
	(5300,1678)(5349,1683)(5397,1688)
	(5444,1693)(5489,1698)(5534,1703)
	(5578,1708)(5620,1713)(5662,1718)
	(5702,1723)(5741,1728)(5779,1734)
	(5817,1739)(5853,1745)(5888,1750)
	(5923,1756)(5957,1761)(5989,1767)
	(5990,1767)(6034,1775)(6078,1783)
	(6120,1792)(6162,1801)(6202,1811)
	(6242,1820)(6281,1831)(6320,1842)
	(6358,1853)(6395,1865)(6431,1878)
	(6467,1891)(6502,1905)(6537,1919)
	(6570,1935)(6603,1951)(6635,1967)
	(6666,1984)(6696,2002)(6725,2021)
	(6754,2040)(6781,2060)(6808,2080)
	(6833,2101)(6858,2123)(6881,2145)
	(6904,2168)(6926,2191)(6947,2214)
	(6967,2239)(6987,2263)(7006,2289)
	(7024,2315)(7042,2341)(7060,2369)
	(7077,2397)(7092,2423)(7108,2450)
	(7123,2478)(7138,2507)(7153,2538)
	(7168,2570)(7184,2603)(7199,2638)
	(7215,2675)(7231,2714)(7248,2755)
	(7265,2798)(7282,2843)(7300,2891)
	(7319,2941)(7338,2994)(7358,3049)
	(7378,3106)(7398,3165)(7419,3226)
	(7440,3289)(7461,3352)(7482,3415)
	(7503,3478)(7523,3540)(7543,3600)
	(7562,3658)(7579,3712)(7595,3762)
	(7610,3806)(7622,3846)(7633,3880)
	(7642,3908)(7649,3930)(7654,3947)
	(7658,3959)(7660,3966)(7661,3970)(7662,3972)
\put(4197,3747){\makebox(0,0)[lb]{\smash{{{\SetFigFont{12}{34.8}{\rmdefault}{\mddefault}{\updefault}$M$}}}}}
\put(7492,107){\makebox(0,0)[lb]{\smash{{{\SetFigFont{12}{34.8}{\rmdefault}{\mddefault}{\updefault}$\phi_0$}}}}}
\end{picture}
}
\caption{Qualitative visualization of the D1 brane profile $M({\phi_0})$
for $\de>0$. }\label{profile2}
\end{figure}
\subsection{Boundary fields}

It is sometimes useful to observe (see appendix \ref{ABCFT} for more details)
that fields $\Phi^\de_{\al}(\si)$ localized on the boundary can be defined
with the help of
\begin{equation}\label{bulkbd}
\Phi^\de_{P}(\si)\,|\,B_{\de}^{}\,\lket\,\equiv\,
C_\de(P)\lim_{\tau\ra 0} \,(2\tau)^{2\De_\al-\De_{2\al}}\,
V_{\al}(\tau,\si)\,|\,B_{\de}^{}\,\lket\,,
\end{equation}
where $P$ and $\al$ are related by $\al=1+iP$.
$C_\de(P)$ is a certain normalizing factor defined
in appendix A
that we do not need explicitly except at the point $C_\de(i\de)=0$.
It has been chosen in such a way that the boundary
fields  $\Phi^\de_P(\si)$ are symmetric under $P\ra -P$,
$\Phi^\de_P(\si)=\Phi^\de_{-P}(\si)$. It will later be important for us
to observe
that the asymptotics of $\Phi^{\de}_{i\varpi}(x)$  for
$\phi_0\ra -\infty$ can be represented as follows
\begin{equation}\label{bdasy}
\Phi^{\de}_{i\varpi}(x)\,\underset{\phi_0\ra-\infty}{\sim}\,
e^{\phi_0}\,\big(\,C_{\de}(i\varpi)\!:\!e^{-\varpi\phi}\!:\!+
C_{\de}(-i\varpi)\!:\!e^{+\varpi\phi}\!:\big)
\end{equation}
The first term is directly understood from
$V_{\al}(\tau,\si)\sim
:\!e^{2\al\phi}(\tau,\si)\!:$, the second is found taking into account that
$\Phi^\de_P(\si)=\Phi^\de_{-P}(\si)$.

In order to characterize the
boundary fields $\Phi^{\de}_P(x)$
completely we need to know both the
operator product expansions (OPEs) and the three point functions
of these fields. The OPE is known \cite{T2} to be of the form
\begin{align}\label{OPE}
\Phi^{{\de}}_{P_\2}(x_\2)
\Phi^{{\de}}_{P_\1}(x_\1)
\;=\;
\int_{0}^{\infty} dP_\3\;  F^{P_\3}_{P_\2 P_\1}\,
& |x_\2-x_\1|^{\De_{P_\3}-\De_{P_\2}-\De_{P_\1}}\,
\Phi^{{\de}}_{P_\3}(x_\1) &
\\
+  f^{\vartheta}_{P_\2P_\1}
 & |x_\2-x_\1|^{\De_{\vartheta}-\De_{P_\2}-\De_{P_\1}}
\Phi_{\vartheta}^{{\de}}(x_\1)
+\;{\rm descendants}\,.
\nn\end{align}
Of particular importance for us is the term in the second line
of \rf{OPE} which is proportional to the
field $\Phi_{\vartheta}^{{\de}}(x_\1)$, with $\vartheta$ being defined
as
\begin{equation}
\vartheta\,=\,i\de\,.
\end{equation}
The field $\Phi_{\vartheta}^{{\de}}(x_\1)$ has
conformal dimension $\De_\vartheta=1-\de^2<1$. It will therefore
correspond to an open string
tachyon in the $c=1$ noncritical string theory  that we are about to
study. The fact that $\Phi_{\vartheta}^{{\de}}(x_\1)$
appears discretely in \rf{OPE} indicates that
it creates a bound
state in the spectrum of boundary Liouville theory on the strip,
as will be further discussed in subsection \ref{BLstrip} below.
The OPE coefficient $f^{\vartheta}_{P_\2P_\1}$
is nonvanishing only if $\de>0$.

We will need the explicit formula for the OPE coefficients
$F^{P_\3}_{P_\2 P_\1}$ 
only in the case when
$\de\ra 0$
with $P_k=\CO(\de)$,
$k=1,2,3$,
\begin{equation}\label{Fapprox}
F_{P_\2P_\1}^{P_\3}\,\dasym\,
\frac{1}{2\pi}\frac{4P^2_\3}{\de^2+P^2_\3}\,.
\end{equation}
This formula is proven in the Appendix \ref{ABCFT}.
As noted above, we have $f^{\vartheta}_{P_\2P_\1}=0$ for $\de<0$.
The OPE is then defined for $\de>0$ by analytic continuation
with respect to the parameter $\de$, see \cite{T2} for more details.
One picks up extra contributions from poles of $F_{P_\2P_\1}^{P_\3}$
which cross the contour of integration in \rf{OPE}.
It then follows easily from \rf{Fapprox}
\begin{equation}\label{fapprox}
f_{P_\2P_\1}^{\vartheta}\,=\,-2\pi i
\Res_{P_\3=i\de} \,F_{P_\2P_\1}^{P_\3}\,
\dasym\, 2\delta\,.
\end{equation}
Note that the formulae \rf{Fapprox} and
\rf{fapprox} also cover the cases
when $P_\2$ or $P_\1$ take the value $\vartheta$.


We will also need the bulk-boundary structure function
$\bra \,P_\1\,|\,\Phi_{P_\2}^{{\de}}(1)\,|\,B_\de^{}\,\ket^{}_{\rm L}
$. It is shown in Appendix \ref{ABCFT} that
in the limit $\de\ra 0$,
$P_\2=\CO(\delta)$ we find the following behavior
\begin{equation}\label{BBd}
\bra \,P_\1\,|\,\Phi_{P_\2}^{{\de}}(1)\,|\,B_\de^{}\,\ket^{}_{\rm L}\,\dasym
\,\frac{1}{{2\pi}}\,
\frac{2\pi P_\1}{\sinh 2\pi P_\1}\,\Theta(P_\1)\,,
\end{equation}
which is {\it independent} of $P_\2$.

\subsection{Hamiltonian picture - open string channel}\label{BLstrip}

There is an alternative Hamiltonian representation which is associated to
the world-sheet being the strip $\BS$.
The corresponding Hilbert space of the boundary Liouville theory
with boundary condition parametrized by $\de$ on both sides of the
strip may then be represented as follows \cite{T1,T2}
\begin{equation}\label{openspec}
\CH_{{\de}{\de}}^{\rm B}\;=\;
\int_{\BR_+}^{\oplus}dP\; \CV_P\,\oplus\,\left\{
\begin{aligned}
\emptyset\;\;& {\rm for}\;\;\de<0\,,\\
\CV_{\vartheta}\;\;& {\rm for}\;\;\de>0\,,
\end{aligned}
\right.
\end{equation}
where $\CV_P$ is the irreducible unitary representation
of the Virasoro algebra with $c=25$ which has highest weight $\De_P=1+P^2$.
State-operator correspondence therefore yields the usual relation between
the  OPE \rf{OPE} of boundary fields to a summation over
a basis for the Hilbert space $\CH_{{\de}{\de}}^{\rm B}$.

We want to show that
the additional contributions which occur in the spectrum \rf{openspec}
can be interpreted as bound states. This can be seen more
clearly by representing the states in terms of the Schr\"odinger
representation for the zero mode $\phi_0^{\rm\sst op}\equiv
\int_{0}^{\pi}d\si \,\phi(\si,\tau)\big|_{\tau =0\,.}$ which is
naturally associated to the given Hamiltonian picture.
The eigenstates of the Hamitonian ${\mathbf H}$  are then
represented by wave-functions $\Psi_P({\phi_0^{\rm\sst op}})$
which have an
asymptotic behavior for ${\phi_0^{\rm\sst op}}\ra-\infty$ of the following form
\begin{equation}
\Psi_P({\phi_0^{\rm\sst op}})\,
\underset{{\phi_0^{\rm\sst op}}\ra -\infty}{\sim}\,\,
\left(\,C_{\de}(P)e^{iP{\phi_0^{\rm\sst op}}}+C_{\de}(-P)
e^{-iP{\phi_0^{\rm\sst op}}}\right)\Om\,
\end{equation}
where $\Om$ is the Fock vacuum.
Our previous observation that $C_{\de}(\vartheta)=0$ therefore
implies the
exponential decay of $\Psi_\vartheta({\phi_0^{\rm\sst op}})$ for
${\phi_0^{\rm\sst op}}\ra\-\infty$
characteristic for a bound state.

In the context of the $c=1$ noncritical string theory there
will be a single physical state $|\vartheta\kket$ which is
constructed by tensoring the highest weight state in
$\CV_{\vartheta}$ with a suitable ``dressing'' from the CFT
associated to the time direction $X_0$.
From the string theoretical point of view it therefore
seems natural to interpret the concentration of ``mass''
as depicted in figure  \ref{profile2} as a potential sink
for the open strings on the D1 branes. The attractive force
associated with the potential sink may bind
open strings.


\subsection{Perturbed boundary Liouville theory} \label{pert_bL}

We will consider the perturbation of boundary Liouville theory which
corresponds to  the boundary action
\begin{equation}\label{Lbdact}
S_{\rm Bd}\,=\,\la\,\int_{\pa\Sigma}dx\;\Phi^{\de}_{\vartheta}(x)
\end{equation}
The perturbed boundary state is then formally defined as
\begin{equation}
|\,B_{\de}^{}\,\ket^{\rm pert}_{\rm L}\,=\,
e^{-S_{\rm Bd}}\,|\,B_{\de}^{}\,\ket^{}_{\rm L}\,.
\end{equation}
Our aim is to extract the leading behavior of the perturbed boundary
state $|\,B_{\de}^{}\,\ket^{\rm pert}_{\rm L}$ for $\de\ra 0$.
As discussed in section \ref{summingRG} we may use
the renormalization group to resum the relevant contributions
of the perturbative expansion in powers of $\la$ into
renormalized couplings.

\subsubsection{RG flow equations}\label{Liouflow}

In order to apply the discussion from section \ref{summingRG} to the
case at hand let us introduce a proper regularization
scheme with cut-off $\ep$ and consider the boundary
action
\begin{equation}
S_{\rm Bd}^{\rm ren}\,=\,\int_{\pa\Sigma}
dx\;\left(\,u\,\ep^{-\de^2} \Phi^{\de}_{\vartheta}(x)+
\int_0^\infty dP\;\la(P)\,\ep^{P^2} \Phi^{\de}_{P}(x)\right)\,,
\end{equation}
which contains the renormalized
coupling constants $u\equiv u_\ep$ and
$\la(P)\equiv\la_\ep(P)$. We have included contributions containing the
irrelevant fields $\Phi^{\de}_{P}(x)$ since fields with $P=\CO(\de)$
are nearly marginal.
Independence of the correlation functions from the cut-off $\epsilon$
follows if the
coupling constants  satisfy
the RG flow equations:
\begin{align}\label{RGlambda}
&\ep\frac{d}{d\ep}u -\de^2u \,=\, -f_{\vartheta\vartheta}^{\vartheta}u^2
-2\int\limits_0^{\infty}dP_\1\;f_{\vartheta P_\1}^{\vartheta}\la(P_\1)u-
\int\limits_0^{\infty} dP_\2dP_\1\;
f_{P_2 P_1}^{\vartheta}\;\la(P_\1)\la(P_\2)\\
&\ep\frac{d}{d\ep}\la(P) +P^2\la(P) \,=\,
- F_{\vartheta\vartheta}^{P}u^2
-2\int\limits_0^{\infty}dP_\1\;F_{\vartheta P_\1}^{P}\la(P_\1)u-
\int\limits_0^{\infty} dP_\2dP_\1\;
F_{P_2 P_1}^{P}\;\la(P_\1)\la(P_\2)
\nn\end{align}
Let us  analyze the equations at or near the new fixed point.
Equations \rf{RGlambda}, \rf{Fapprox}, \rf{fapprox} then suggest that $\la(P)$ must be peaked
for small $P$.
Plugging in the approximate formulae \rf{Fapprox}, \rf{fapprox} one simplifies
the RG flow equations to
\begin{align}
\ep\frac{d}{d\ep}u & =\delta^2 u -2\delta w^2\,, \\
\ep\frac{d}{d\ep}\la(P) & =-P^2\la(P)
-\frac{1}{2\pi}\frac{4P^2}{\de^2+P^2}w^2\,,\label{RGsimple}\end{align}
where
\begin{equation}\label{vdef}
w\equiv u+\int_0^{\infty} dP\;\la(P)\;.
\end{equation}
We may observe that the RG flow equations have a
fixed point $\la_*(P)$.
First note that one can determine the
$P$-dependence of $\la_*(P)$ for $P\in\BR_+$
from \rf{RGsimple}  and \rf{vdef}
to be
\begin{equation} \label{fpt}
\la_*(P)\;=\;\,\frac{2}{\pi}\,\frac{\de\,v_*}{\de^2+P^2}\,,\qquad
v_*\,\equiv\,\int_0^{\infty}dP \;\la_*(P)\,.
\end{equation}
Inserting this into \rf{RGsimple} yields the following
equations for the fixed point values $u_*$
and $v_*$:
\begin{align}
0&=\delta u_* -2(u_*+v_*)^2\,, \\
0&=-\delta v_* -(u_*+v_*)^2\,.
\end{align}
This implies in particular that
\begin{equation}\label{fp_vals}
u_{*}=2\delta\, , \quad v_{*}=-\delta\, , \quad w_*\,=\,u_*+v_*\,=\,\de\,,
\end{equation}
are the coupling constants at the
new fixed point of the renormalization group.

\subsubsection{Determination of the new fixed point}

We will next determine the perturbed boundary state $|B_*\ket$
at the new RG
fixed point. We claim that
\begin{equation}\label{newBd}
|\,B_*\,\ket^{\rm pert}_{\rm L}\;=\;
|\,B_{-\de}^{}\,\ket^{}_{\rm L}
\,.
\end{equation}
In order to verify \rf{newBd} let us first note that
the new fixed point must have the same value of $\mub$.
This can be seen as follows. We had previously seen that
the wave-function of the
boundary state $B_\de({\phi_0})$ approaches a constant for
$\phi_0\ra-\infty$. Subleading contributions
for $\phi_0\ra-\infty$ that decay exponentially will
produce poles in amplitudes such as
$A^\de_P=\bra P | B_{\de}^{}\lket$ which come from the asymptotic behavior
of the integrand in \rf{bdoverlap}. The next-to leading order contribution
to $B_\de({\phi_0})$ is of the order $e^{\phi_0}$ which
corresponds to the pole of $A^\de_P$ at $P=-i/2$, see \rf{onept}.
It is important
to note that this contribution is proportional to $\mub$, as may also be
inferred from \rf{onept}. Note, however, that the perturbing boundary field
$\Phi^{\de}_{\vartheta}(x)$ behaves asymptotically as
$e^{(1+\de)\phi_0}$. Indeed, a generic boundary field
$\Phi^{\de}_{i\varpi}(x)$ has asymptotic behavior of the form given
in equation \rf{bdasy}. The first term in \rf{bdasy} vanishes in the
case of the perturbing boundary field $\Phi^{\de}_{\vartheta}(x)$ as follows
from $C_\de(i\de)=0$. This means that the perturbative expansion
of $|B_{\de}^{}\ket^{\rm pert}_{\rm L}$ in powers of $\la$
will only generate
terms which vanish faster than $e^{\phi_0}$ when $\phi_0\ra\infty$.
The value of $\mub$ must therefore be unchanged.

In order to further
confirm our prediction \rf{newBd}
let us calculate, to lowest order in $\de$, the
deviation $d_*(P)$:
\begin{equation}
d_*(P)\,=\,\bra \,P\,|\,B_*\,\ket^{}_{\rm L}-
\bra \,P\,|\,B_\de^{}\,\ket^{}_{\rm L}\,.
\end{equation}
We find
\begin{align}
d_*(P)\,\dasym\, &\;-2\pi \,u_*\,
\bra \,P\,|\,\Phi_{\vartheta}^{{\de}}(0)\,|\,B_\de^{}\,\ket^{}_{\rm L}
\nn\\
& \,-2\pi \int_0^{\infty}dP'\;\la_*(P')\,
\bra \,P\,|\,\Phi_{P'}^{{\de}}(0)\,|\,B_\de^{}\,\ket^{}_{\rm L}
\,.\label{d*def}
\end{align}
The relevant bulk-boundary
correlation functions are given in equation \rf{BBd}. Note that
they are {\it independent} of $P'$.
We may therefore carry out the integral
over $P'$  in \rf{d*def} by using the definition of $v_*$
given in \rf{fpt}.
It follows that
\begin{equation}
d_*(P)\,\dasym\, -2\pi \,w_*\,
\bra \,P\,|\,\Phi_{\vartheta}^{{\de}}(0)\,|\,B_\de^{}\,\ket^{}_{\rm L}
\,.
\end{equation}
By using equations \rf{FP} and \rf{BBd} we arrive at
the following formula for the perturbed one point function
at the new fixed point:
\begin{equation}\label{pertonept}
d_*(P)\,\dasym\,-\delta\,\frac{2\pi P}{\sinh 2\pi P}\Theta(P)\,.
\end{equation}
This is the same result as one would have found from
\rf{newBd} by keeping terms up to $\CO(\de)$.

The time-independent treatment described in this section may lead one
to conjecture that there should exist a time-dependent solution
of string theory which interpolates between
asymptotic states that are D1 branes with
labels $\de$ and $-\de$ respectively. This is what we are going to
construct in the following section.


\section{A noncritical time-dependent string background}\label{background}
\setcounter{equation}{0}


We now want to apply the formalism developed in section \ref{TsummingRG}
to the case that the ${\rm CFT}_{\rm S}$,
the conformal field theory which describes the spatial part
of the background is (boundary) Liouville theory
with $c=25$. The boundary action which will  define the perturbed
time-dependent background is then given by the expression
\begin{equation}\label{ALbare}
  S_{\rm Bd}\,\equiv\,\la \,\int_{\pa\Sigma}dx\; [e^{\de X_0}\Phi^\de_{\vartheta}](x)\,. \end{equation}
Note
that the short distance singularities in
the OPE of $[e^{\de X_0}\Phi_\vartheta^\de](x)$
with itself are integrable.
It follows that \rf{pertB}
 indeed defines a conformal boundary state
to all orders in a formal expansion in the parameter $\la e^{\de t}$.

In order to describe the time evolution of the perturbed
boundary state
$|\,B_{\de}^{}\,\kket_{\rm dyn}^{}$ we will
consider the amplitude
\begin{equation}\label{APtdef}
A(P,t)\,\equiv\,\bbra \,P,t\, |\,B_{\de}^{}\,\kket_{\rm dyn}^{}\,,\qquad \bbra
\,P,t\,|\,\equiv\,\langle P|\ot\xbra{t}\,,
\end{equation}
from which the corresponding amplitude
\begin{equation}\label{APwdef}
A(P,\om)\,\equiv\,
\bbra \,P,\om\, |\,B_{\de}^{}\,\kket_{\rm dyn}^{}\,,\qquad \bbra
\,P,\om\,|\,\equiv\,\langle P\,|\ot\xbra{\om}\,
\end{equation}
will then follow by Fourier transformation.

\subsection{RG improvement in the case of a near-marginal continuum}

In order to employ the technique from section \ref{summingRG} to our specific example let us assume
having introduced a proper  regularization scheme with a short-distance cut-off $\ep$. We will
consequently have to work with a renormalized boundary action density $S_{\rm Bd}^{\rm ren}(x)$
which will be of the form
\begin{align}\label{ABdexp}
S_{\rm Bd}^{\rm ren}(x)&\,=\,\sum_{n=1}^{\infty}\, \Bigl( U_{n} \ep^{(n^2-1)\de^2} [e^{n\delta
X_{0}} \Phi^{\delta}_{\vartheta}](x) + \int\limits_{0}^{\infty}dP\; \la_n(P)
\ep^{n^2 \de^2 + P^2} [e^{n\de X_0} \Phi_P^{{\de}} ](x)\Bigr)\,.
\end{align}
The conditions for the $\ep$-independence of the
correlation functions are then found to be the equations
\begin{align}\label{RGflowtime}
&\ep\frac{d}{d\ep}U_{n} +(n^{2}-1)\de^2U_{n} \,=\\
& =\sum\limits_{m=1}^{n-1}\Bigl( -f_{\vartheta\vartheta}^{\vartheta}U_{m}U_{n-m}
-2\int\limits_0^{\infty}dP_\1\;f_{\vartheta P_\1}^{\vartheta}U_{m}\la_{n-m}(P_\1)-
\int\limits_0^{\infty} dP_\2dP_\1\;
f_{P_2 P_1}^{\vartheta}\;\la_{m}(P_\1)\la_{n-m}(P_\2)\Bigr)\,,\nn \\
\label{RGflowtime2}
&\ep\frac{d}{d\ep}\la_{n}(P) +(n^{2}\delta^{2}+P^2)\la_{n}(P) \,=\\
&=\sum\limits_{m=1}^{n-1}\Bigl( - F_{\vartheta\vartheta}^{P}U_{m}U_{n-m}
-2\int\limits_0^{\infty}dP_\1\;F_{\vartheta P_\1}^{P}U_{m}\la_{n-m}(P_\1)-
\int\limits_0^{\infty} dP_\2dP_\1\; F_{P_2 P_1}^{P}\;\la_{m}(P_\1)\la_{n-m}(P_\2)\Bigr)\,.
\nn\end{align}
Removing the cut-off $\ep$ by sending $\ep\ra 0$ should reproduce the bare action \rf{Abare}. This
means that we are interested in the solution to \rf{RGflowtime} which is defined by the following
supplementary conditions:
\begin{equation}
\label{timeBC}
\begin{aligned}
&\lim_{\ep\ra 0}\,\la_n(P) \ep^{n^2\de^2 + P^2} \,=\,0\;\;{\rm for}\;\;P\in\BR_+\,,\\
&\lim_{\ep\ra 0}\,U_n \ep^{(n^2-1)\de^2} \,=\,0 \;\;{\rm for}\;\;n>1\,,\\
&\lim_{\ep\ra 0}\,U_1 \,=\,\nu\de\, . \end{aligned} \end{equation}
$\lambda\equiv\nu\de$ is
the value of the corresponding ``bare'' coupling introduced in \rf{pertB}. It will again
turn out to be useful to measure $\la$ in units of $\de$ by introducing $\nu=\la/\de$.

 Equations \rf{RGflowtime}, \rf{timeBC} can be solved recursively. It is easy
to see that we must have $U_1=\la$ and $\la_1(P)=0$ for $P\in\BR_+$. For $n>1$ let us note that a
special solution to the inhomogenous equations \rf{RGflowtime} is always given by $\ep$-independent
coupling constants $\la_n(P)$. The general solution of the equations \rf{RGflowtime} is then
obtained by adding an arbitrary solution to the homogeneous equations which are obtained from
\rf{RGflowtime} by dropping the right hand side. However, the solutions to these homogeneous
equations will never satisfy the boundary conditions \rf{timeBC} unless they are identically zero.
We therefore find that $\la_n(P)$ is determined for $n\geq 2$ from the recursion relations
\begin{align}\label{recrel} & (n^2\de^2 +P^2)\la_n(P) \,=\,  \\  \nn
&-\sum\limits_{m=1}^{n-1}\Biggl(
 F_{\vartheta\vartheta}^{P}U_{m}U_{n-m} +
 2\int\limits_0^{\infty}dP_\1\;F_{\vartheta P_\1}^{P}U_{m}\la_{n-m}(P_\1)+
\int\limits_0^{\infty} dP_\2dP_\1\; F_{P_2 P_1}^{P}\;\la_{m}(P_\1)\la_{n-m}(P_\2)\Biggr)\, .
\end{align} It follows that $\la_n(P)\propto (n^2\de^2 +P^2)^{-1}$ is strongly peaked around
$P=\CO(\de)$. Noting that the OPE coefficients $F_{P_2 P_1}^{P}$ are approximately constant in this
region, cf. eqn. \rf{Fapprox}, allows us to get the $\delta\to 0$ asymptotics of   the integrations
over $P_\2$, $P_\1$
by introducing a rescaled momentum variable \begin{equation} q = P/\delta \, \end{equation} and
substituting the asymptotic values (\ref{Fapprox})
at fixed $q_{i}$'s
into (\ref{recrel}). Define
\begin{equation}
V_n\,\equiv\, \int_0^{\infty} dP\;\la_n(P)\,.
\end{equation}
One finds then from (\ref{recrel}), (\ref{Fapprox}) that the $P$-dependence of
 $\la_n(P)$ (at $\delta \to 0$)  
can  be written in the form
\begin{equation}\label{lambda_modes}
{\lambda}_n(P)\;=
\lambda_{n}(q\delta) = \;\frac{V_n}{\delta}\,\frac{n+1}{n^2+q^2}\, \frac{4q^2}{2\pi(1+q^2)}\,.
\end{equation}
 One further finds that (\ref{RGflowtime}) and
\rf{RGflowtime2} lead to
a closed set of equations for the coupling constants $V_{n}$ and $U_{n}$,
\begin{equation}\label{UVrecrel}
\begin{aligned}
 (n^2-1)&\de U_n\,=\,-2R_n\,, \\
(n+1)&\de V_n\,=\,-R_n\,,
\end{aligned}\qquad R_n\,\equiv\,
\sum_{m=1}^{n-1}\Big(U_mU_{n-m}+2U_mV_{n-m}+V_{m}V_{n-m}\Big)\,.
\end{equation}
These recursion relations combined with the initial conditions
\begin{equation}\label{inicond}
U_1\,=\,\de\nu,
\qquad
V_1\,=\,0\,,
\end{equation}
completely determine the coupling constants $U_n$ and $\la_n(P)$ via \rf{lambda_modes}.
It is easy to show that \rf{UVrecrel} and \rf{inicond} imply that
$U_n=\CO(\de)$ and $V_n=\CO(P)$. The integral over $P$ in \rf{ABdexp} will also
be of the order $\de$ since $\la_n(P)$ is peaked around $P=\CO(\de)$.
In order to extract the leading behavior of $|\,B_{\de}^{}\,\kket_{\rm dyn}^{}$
for $\de\ra 0$ we may therefore indeed work with the renormalized action
\rf{ABdexp} in the following.

\subsection{Perturbed one point function}

As a simple example for the
application of our findings let us now consider the
amplitude $A(P,t)$. It will be convenient to subtract the constant initial value of
this quantity and consider
\begin{equation}
D(P,t)\,\equiv\, \bbra \,P,t\,
|\,B_{\de}^{}\,\kket_{\rm dyn}^{}-
\bbra \,P,t\,
|\,B_{\de}^{}\,\kket_{\rm stat}^{}\,.
\end{equation}
According to the discussion in our previous subsection we may calculate this quantity
to leading order in $\de$ by expanding \rf{ABdexp} to the first order,
\begin{align}
D(P,t)\,\dasym\, &\;-2\pi \,U(t)\,
\bra \,P\,|\,\Phi_{\vartheta}^{{\de}}(0)\,|\,B_\de^{}\,\ket^{}_{\rm L}
\nn\\
& \,-2\pi \int_0^{\infty}dP'\;\la_{P'}(t)\,
\bra \,P\,|\,\Phi_{P'}^{{\de}}(0)\,|\,B_\de^{}\,\ket^{}_{\rm L}
\,,\label{DDef}
\end{align}
where we have used the notations
\begin{equation}\label{functions}
\lambda_P(t)=\sum_{n=1}^{\infty}e^{\de nt}{\lambda}_n(P)\,,
\qquad U(t)=\sum_{n=1}^{\infty}e^{\de n t}U_n\,,
\end{equation}
The relevant correlators are given in equation \rf{BBd}. Note that
they are {\it independent} of $P'$.
We may therefore carry out the integral
over $P'$  in \rf{DDef} as follows:
\begin{equation}\begin{aligned}
\int_{0}^{\infty}dP'\;\la_{P'}(t)\,=\,\sum_{n=1}^{\infty}e^{\de nt}
\int_{0}^{\infty}dP'\;\la_n(P')\,=\,\sum_{n=1}^{\infty}e^{\de nt}V_n\,\equiv\,
V(t)\,.
\end{aligned}
\end{equation}
Inserting the explicit expression \rf{BBd} leads to
\begin{equation}\label{DPtexpl}
D(P,t)\,\dasym\, -W(t)\,\frac{2\pi P}{\sinh 2\pi P}\Theta(P)\,,\quad W(t)\equiv U(t)+V(t)\,.
\end{equation}
The task remains to calculate $W(t)$ explicitly.
Note that so far we had to assume that $t$ is sufficiently small. We will indeed see
that the range of convergence of the series \rf{functions} is finite. However,
the function $W(t)$ will turn out to have an analytic continuation which allows us to
extend the definition of $D(P,t)$
to all real values of $t$.


\subsection{Time evolution}

The recursion relations \rf{UVrecrel} are easily translated into differential equations
for the generating functions \rf{functions},
\begin{equation}\label{ev}
\begin{aligned}
& \de^{-1}\ddot{U}=\delta U-2 W^2,\\
& \dot{V}=-\delta V-W^2, \end{aligned}\qquad W\equiv U+V\,. \end{equation}
The dots indicate
derivative with respect to $t$.
In this section we will find the
explicit solutions to the time-evolution equations (\ref{ev}) with the boundary conditions \rf{inicond}.
This will then
allow us, using (\ref{lambda_modes}), to find the function $\lambda_{P}(t)$.

To begin with, let us note that the time evolution equations (\ref{ev}) can be rewritten in the
form
\begin{equation} \ddot U = -2\delta \frac{\partial \CP}{\partial U} \,,
\qquad \dot V = - \frac{\partial \CP}{\partial V} \label{ev2}\,, \end{equation}
where \begin{equation}\label{P} \CP=\CP(U,V)=
-\frac{\delta}{4}U^{2} + \frac{\delta}{2}V^{2} +
\frac{1}{3}(U+V)^{3} \, . \end{equation}
The existence of such a ``potential'' function $\CP(U,V)$
can be traced back to the cyclic symmetry of the OPE coefficients.  However equations (\ref{ev2})
are  clearly not the standard Euler-Lagrange equations for two mechanical degrees of freedom due to
the first order derivative term $\dot V$. The dissipative nature of this system of equations can be
discerned by looking at the time derivative \begin{equation} \frac{d}{dt}\left[\frac{1}{2}\dot
U^{2} + \CP(U,V)\right]
\,=\, -(\dot V)^{2} \,\le\, 0 \, . \end{equation} Here in the square brackets
we have an expression that can be thought of as an effective energy for the $U$-degree of freedom
which ought to monotonically decrease due to this identity.

Although as we have just demonstrated the energy is no longer an integral of motion for equations
(\ref{ev2}), there is  another integral of motion. One can easily show that \begin{equation}
\label{iom} \frac{d}{dt}\left[e^{\delta t}(\delta^{-1}\dot U - U - 2V)\right] \,=\, 0 \, .
\end{equation}

Using the initial conditions at $t\to -\infty$, that are essentially given by the fact that our
solutions in that limit can be represented as power series (\ref{functions}), we see that the
integral of motion in \rf{iom}  assumes zero value. This reduces (\ref{ev2}) to a system of first
order differential equations \begin{eqnarray}
&&\dot U = \delta(U + 2V)\nonumber \\
&&\dot V = -\delta V - (U+V)^{2} \label{ev3}\,. \end{eqnarray} One further observes that the
$W=U+V$ degree of freedom decouples: \begin{equation} \dot W = \delta W - W^{2} \, . \end{equation}
Solving this equation we obtain
\begin{equation}\label{Wt}
W(t)\, =\, \frac{\delta e^{\de t}}{C+e^{\delta t}}\,,
\end{equation}
where $C$ is a constant of integration.
Substituting (\ref{Wt})  back into (\ref{ev3}) we obtain
\begin{align}\label{Uoft} & U(t) =
2\delta[1 - Ce^{-\delta t}\ln(1 + C^{-1}e^{\delta t})] = \delta
\sum\limits_{n=1}^{\infty}(-1)^{n+1}e^{n\delta t}C^{-n} \, , \\
& \label{Vn} V(t) = W(t) - U(t) = \delta \sum_{n=2}^{\infty}(-1)^{n}\left(
\frac{1-n}{1+n}\right) C^{-n}e^{n\delta t}
\end{align}
The integration constant $C$ is found to be related to the bare
coupling $\la=\nu\de$ by observing that
(\ref{Uoft}) implies $U_{1}=\delta C^{-1}$.
Taking into account the condition (\ref{timeBC}) therefore yields the relation
\begin{equation}
C\,=\,\delta\lambda^{-1}
=\,\nu^{-1}\,.
\end{equation}

We are left with the task to calculate $\la_P(t)$ explicitly.
Plugging the coefficients $V_{n}$ from (\ref{Vn}) into (\ref{lambda_modes}) we obtain
\begin{align} \label{l1}
&\lambda(q, t)  \equiv \lambda_{q\delta}(t)=
\sum_{n=1}^{\infty}
\lambda_{n}(q\delta)e^{n\delta t} =
\frac{2q^{2}}{\pi(1+q^{2})}\tilde \lambda(q,t)\, , \nonumber \\
& \tilde \lambda(q,t) = \sum_{n=2}^{\infty}\frac{1-n}{n^{2} + q^{2}}(-\nu e^{\delta t})^{n} \, .
\end{align}
The function $\tilde \lambda_P(t)$ can be represented as
\begin{equation}\label{l2}
\tilde \lambda(q,t) = f_q(t) - \delta^{-1}\frac{d}{d t} f_q(t)
\end{equation}
where
\begin{align}\label{l3}
 f_q(t) =
-\frac{\nu e^{\delta t}}{2q(q^{2}+1)}\Bigl[ & (q-i)\,
{}_{2}F_{1}(1-iq,1; 2-iq; -\nu e^{\delta t})
+ \nonumber \\ +
& (q+i)\, {}_{2}\!\,F_{1}(1+iq, 1; 2 +iq; -\nu e^{\delta t})\Bigr] \, .
\end{align} Formulas
(\ref{l1}), (\ref{l2}), (\ref{l3}) provide  explicit expressions for the time-dependent couplings
$\lambda_P(t)$ via hypergeometric functions.

\subsection{Asymptotics $t\ra\infty$}

Having found the explicit expression \rf{Wt} for the function $W(t)$ we may now return
to the discussion of the perturbed one point function $A(P,t)$. Observing that the expression
\rf{Wt} is well-defined for all values of $t$ will
not only allow us to to discuss the
asymptotics of $A(P,t)$
for $t\ra\infty$, it will also finally give us the leading
order result for the amplitude $A(P,\om)$.

To begin with, let us observe that the asymptotic values
\begin{equation}\label{FP}
\lim\limits_{t\to \infty} U(t) =2\delta=u_* \, , \quad
\lim\limits_{t\to \infty} V(t) =-\delta=v_*\,,\quad
\lim_{t\ra\infty}W(t)\,=\,\delta =w_*
\, ,
\end{equation}
coincide with the fixed point values $u_*$, $v_*$ and $w_*$ that we had found
in the time-independent treatment of subsection \rf{Liouflow}.
The time evolution of $U$ and $V$ therefore smoothly
interpolates between the values of the
corresponding couplings at
the UV and IR fixed points in the time-independent picture.
The corresponding asymptotic values of $A(P,t)$ can be identified with
the overlaps $\bra P |B_{\de}^{}\lket$ and
$\bra P|B_{*}^{}\lket$, respectively.
Let us finally remark that the values $u_{*}$, $v_{*}$ correspond to a
local minimum of the ``potential'' function $\CP$ (\ref{P}).

It will also be quite suggestive to look at the asymptotics of $\lambda(q,t)$
for  $t\to \infty$. It can be deduced from the asymptotic of the hypergeometric function
for $|x|\ra \infty$,
\begin{equation}
{}_{2}\!\,F_{1}(1-iq,1;2-iq;x)\,\underset{|x|\ra\infty}{\to}\, (-x)^{-1}\frac{q+i}{q} +
(-x)^{-1-iq}\Gamma(2-iq)\Gamma(iq) \,.
\end{equation}
Using this asymptotics we obtain
\begin{equation} \label{rad}
\lambda(q,t) \to
-\frac{2}{\pi(1+q^{2})}  +\frac{1}{\sinh(\pi q)}\Bigl( e^{iq\delta t}\nu^{iq}\frac{q}{1+iq} +
e^{-iq\delta t}\nu^{-iq}\frac{q}{1-iq}\Bigr) \,.
\end{equation}
The first term, which is time
independent, coincides with
the fixed point function $\lambda_{*}(P)$
given in (\ref{fpt}), (\ref{fp_vals}).
Together with the asymptotic values (\ref{FP})
this implies that the stationary part of the
$t\to \infty$ asymptotics of the coupling constants corresponds
to the fixed point found in subsection
\ref{pert_bL} which we had
identified with the D1 branes labelled by the
parameter  $-\delta$. The
oscillatory part of  (\ref{rad}) corresponds to a
perturbation of this fixed point background by
marginal operators $\Phi_{P}^{\delta}e^{\pm i PX_{0}}$
with appropriate coupling constants. This
time-dependent piece can be interpreted as open string  radiation on top of
the D1 brane with label $-\delta$.

In order to round off the discussion let us finally calculate
$D(P,\om)\equiv \bbra P,\om
|B_{\de}^{}\kket_{\rm dyn}^{}-
\bbra P,\om
|B_{\de}^{}\kket_{\rm stat}^{}$.
We simply have to perform the inverse Fourier transformation from $\xbra{t}$
to $\xbra{\om}$ by using
\begin{equation}
\widehat{W}(\om)\,\equiv\,
\int_{\BR}\frac{dt}{2\pi}\;e^{i\om t}\,W(t)\,=\,
\frac{i\nu^{-i\frac{\om}{\de}}}{2\sinh\left(\pi \frac{\om}{\de} \right)}\,.
\end{equation}
We arrive at the expression
\begin{equation}\label{DPomexpl}
D(P,\om)\,\dasym\, -
\frac{i\nu^{-i\frac{\om}{\de}}}{2\sinh\left(\pi \frac{\om}{\de}\right)}
\frac{2\pi P}{\sinh 2\pi P}\Theta(P)\,,
\end{equation}
from which the closed string emission in the decay of a D1 brane with
parameter $\de$ into a D1 brane with parameter $-\de$ can be
calculated.


\section{Discussion}
In this paper we analyzed a particular model of D1-brane decay in non-critical $c=1$
string theory. The presence of a small parameter $\delta$ responsible for the mass of the
tachyon allowed us to analyze quantitatively some features
of   the time-dependent CFT that describes the tachyon condensation.
 In particular, employing the RG-resummation technique,
   we found the boundary state for the model in the leading order in the $\delta$-expansion.
   We could show  that in the $t\to \infty$ limit the time-dependent CFT looks like a
  certain  static background describing another D1-brane perturbed by  a time-dependent
marginal perturbation describing an open string radiation propagating to infinity.
We showed that the static D1-brane background at hand coincides with the end point of the RG evolution
 triggered by perturbing the Liouville part of the theory  by the relevant operator
corresponding to the tachyon.

The issue  of what is the relation between the RG flow triggered by a relevant operator
 corresponding to the tachyon
and the description of its condensation by a time-dependent CFT
 was recently addressed in \cite{FHL} in the framework of closed string theory.
The RG-equations are first order in the RG ``time'' and have a dissipative nature as
demonstrated in general by Zamolodchikov's $c$-theorem. On the other hand the
time-evolution equations in  target space are (at least quasiclassically)
second order in time and, superficially, preserve the total space-time energy.
This seems naively to preclude any simple qualitative relationship between the two.
It was shown however in \cite{FHL} that if one properly accounts for dilaton couplings
a damping force appears for the time-evolution equations for the rest of the couplings.
The last feature makes it in principle possible for the time evolution to end in an
RG fixed point accompanied by a time-dependent dilaton and some  examples of this
situation were discussed in \cite{FHL}.

In the case of open string condensation no time-dependent dilaton couplings enter the consideration
 at least at tree level. For the model studied in the present paper  a different type
 of relationship with RG flows  was found.
 In this case the time-dependent CFT describes a localized tachyonic degree of freedom interacting
 with a continuum of open string scattering states. With the total energy being preserved
 the tachyonic degree of freedom relaxes asymptotically into the RG fixed point at the
 expense of producing open string radiation that escapes to infinity.

It seems natural to expect that generalizations of this mechanism
will be applicable in a wide range of situations where
tachyons are localized in a non-compact target space and the boundary conditions
at infinity do not change in the course of tachyon condensation. In the present
case the dominant decay channel was open string radiation, but
in other cases like the example discussed in \cite{T2} one will find
that most of the energy is carried away by closed string radiation.

For the cases in which the tachyon
condensation produces changes of the boundary conditions at infinity
the world sheet description is expected to be more subtle
(see \cite{SeibSh} for a general discussion and \cite{T2} for
some related observations concerning time-dependent phenomena). Reaching the new fixed point requires
giving an expectation value to a non-normalizable operator that seems to be hard to generate
(at least perturbatively) from localized tachyons that are normalizable modes.
One should also note that the very notion of radiation becomes problematic in such cases.
In the case of the model of unstable D1 branes studied in this paper the new fixed point
has the same value of $\mu_{B}$ and thus does not involve changing the boundary conditions
at infinity \cite{T2}.

  It may happen that an unstable system of D-branes decays into a state not carrying
any open string scattering states. In that case closed string radiation may be the dominant mode of decay.
For the present model the closed string radiation appears as a subdominant effect proportional to
the string coupling constant. It would be interesting however to study  back reaction effects on the open string
tachyon condensation due to the closed string radiation for the present model. We leave this issue
for future work.

Another question that is worth  clarification is
a precise relation between the oscillatory  piece of the asymptotic (\ref{rad} )
and open string pair creation as may be measured by a suitable two-point function. This will
require  constructing  marginal operators in the time-dependent theory and finding   their
decomposition into in-coming and out-going scattering states.
We are planning to address these questions in  future work.

At the technical level the present model may look very similar to localized closed string tachyons
in ${\mathbb C}/{\mathbb Z}_{N}$ non-supersymmetric orbifolds of critical superstrings (see
\cite{APS} for the initial discussion)
for large values of $N$.
The tachyon in those models lives in the twisted sector and describes a localized degree of freedom.
Its mass goes to zero when $N$ becomes very large so it looks like one has at his/her disposal
a small parameter similar to $\delta$ in the present model. It may therefore look appealing to
apply methods of the present paper to those models in the $N\to \infty$ limit.
There is however an important distinction between the ${\mathbb C}/{\mathbb Z}_{N}$ theories and models analogous to the one studied in this paper.  The tachyon condensation
in those models does involve changing boundary conditions at infinity.
Furthermore the world sheet analysis of \cite{SarSat}, \cite{DIR} seems to indicate
that the final fixed point for the RG flow is
not  reachable through the $1/N$ expansion.
This unpalatable feature may also be linked to the change of boundary conditions at infinity.
It seems to be important to understand  better how to handle these models   from the world
sheet point of view.



\begin{center}
{\bf Acknowledgments}
\end{center}
A.~K. and J.~T. would like to thank Kavli Institute for Theoretical Physics, where their collaboration
on this project started,
for warm hospitality. Part of this work was done during visits of K.~G. and
A.~K.  to DESY Theory Group whose hospitality  is gratefully acknowledged.
A.K. wants to thank Davit Sahakyan for useful discussions.
The work of A.K. was supported in part by DOE grant DE-FG02-96ER40949.
The work of K.G. was supported by the EUCLID Network, contact number HPRN-CT-2002-00325.

\newpage
\appendix
\setcounter{equation}{0}

\section{Perturbations with UV divergences}

To simplify the discussion in sections \ref{summingRG}
and \ref{TsummingRG} we had been assuming the
absence of UV divergencies. However, it will not be hard
to show that the main results carry over to the case when
there are UV divergencies which arise from the
presence of the identity field in some operator product
expansions.

\subsection{Comments on the time-independent case}

As a typical example of where UV divergences would appear,
consider a CFT with discrete spectrum and let our near marginal field fuse with itself to give the identity,
\begin{align}
  \phi_0(\tau) \phi_0(0) = C_{00}{}^{1\!\!1} \left( 2 \sin{ \tfrac \tau 2 } \right)^{-2+2y_0}
  1 \!\! 1 + \text{other terms} \; .
\end{align}
More generally, we can consider the field fusing to create many {\em non-nearly marginal relevant fields}.
To regulate these divergences one simply repeats the steps of section \ref{sec:RGintro}.
A subtlety comes in specifying renormalisation conditions to replace \eqref{eq:RGcond} which
will no longer make sense in general.  One choice is to introduce a renormalisation scale $\Lambda$
and define physical couplings $\hat{\mu}_k$ via $ \hat{\mu}_k {=} \mu_k(\Lambda) $.
Singular terms in the expression $\mu_k(\ep)$ will then cancel similar singularities in the perturbation
expansion such that the resulting correlation functions are finite and independent of $\ep$.

However, through all this the renormalisation group equations \eqref{eq:RG} are unchanged.
It then follows that if all couplings are assumed to be small and the system flows to a non-trivial fixed point of the RG-equations, the couplings to non-nearly marginal fields will be $\CO(\de^2)$ and so can be ignored to this level in the analysis.  Having reduced the system to that involving only nearly-marginal fields,
the leading $\de$-behaviour of the remaining couplings $ \mu_k(\ep)$ is independent of $\ep$ and the results of the text apply.

In conclusion, even in the presence of UV divergences involving non-nearly marginal fields, the leading behaviour of correlation functions is given
by \eqref{eq:finalren} wherein $\mu_k(\ep)$ can be calculated from the renormalisation group equations involving
only nearly marginal fields.
In the case where the UV divergences come from nearly marginal fields, things are more complicated and we have nothing to say at this time.

\subsection{Time-dependent perturbations in the presence of divergencies}

We would like to convince ourselves that the main features
of the discussion in subsection \ref{TsummingRG2} are still valid
if there are divergencies in the perturbative expansion.

We will assume that the fields $\phi_a$ in the complement $N=F\setminus M$
of the set of all marginal fields $M$ are all such that $y_a=\CO(1)$ when
$\de\ra 0$. We use the letters $a,b,c,\dots$ to label the elements of $N$.
$\de^2$ will be identified with the largest possible
value that $y_i$ takes for $\phi_i\in M$. To simplify life
we will furthermore assume that all relevant OPE coefficients
$C_{ij}^k$, $C_{aj}^k$, $C_{ab}^k$, $C_{ij}^c$, $C_{aj}^c$ and $C_{ab}^c$
are of order $\CO(1)$ when $\de\ra 0$.

We are then led to consider
the regulated
perturbation,
\begin{equation}\label{TSreg}\begin{aligned}
  S_{\text{reg}} \,=\,\sum_{n=1}^\infty\,\Biggl[\,&
\sum_{k \in M}  u_{k,n}
\ep^{n^2\delta^2-y_k} \int dx\;
  [ e^{n \delta X^0} \phi_k ] (x)\\
+ & \sum_{a \in N}v_{a,n}
\ep^{n^2\delta^2-y_a} \int dx\;
 [ e^{n \delta X^0} \phi_a ] (x)\Biggr]\,.
\end{aligned}
\end{equation}
Following the arguments in subsection \ref{TsummingRG2} now leads to
the following recursion relations
\begin{align}\label{Trecrel1}
 &  (y_k- n^2 \delta^2)u_{k,n} = \\
&\quad=  \sum_{m=1}^{n-1} \, \Biggl[\,
\sum_{a,b\in N} C_{ab}{}^k v_{a,m} v_{b,m}+2
\sum_{a\in N}\sum_{i\in M} C_{ai}{}^{k}v_{a,m} u_{i,n-m}+\sum_{i,j\in M}
C_{ij}{}^k u_{i,m} u_{j,n-m}\Biggr]\,,\nn \\
 &  (y_c- n^2 \delta^2)v_{c,n} = \label{Trecrel2}\\
&\quad=  \sum_{m=1}^{n-1} \, \Biggl[\,
\sum_{a,b\in N} C_{ab}{}^c v_{a,m} v_{b,m}+
2\sum_{a\in N}\sum_{i\in M} C_{ai}{}^{c}v_{a,m} u_{i,n-m}+\sum_{i,j\in M}
C_{ij}{}^c u_{i,m} u_{j,n-m}\Biggr]\, .\nn
\end{align}
These equations are supplemented with the boundary
conditions
\begin{equation}
   u_{k,1} = \kappa_k\de^2 \; ,
  \qquad
   u_{k,1} = 0 \; ,
  \qquad v_{a,1}=0\;.
\end{equation}
Keeping in mind that $y_c=\CO(1)$ we may easily
deduce from \rf{Trecrel1}, \rf{Trecrel2} that
$u_{k,n}=\CO(\de^2)$, but $v_n=\CO(\de^4)$. Working to
leading order in $\de^2$ we may therefore simplify
\rf{Trecrel1} to
\begin{equation}
u_{k,n} = \frac 1{y_k- n^2 \delta^2}  \sum_{m=1}^{n-1} \,
\sum_{i,j\in M}
C_{ij}{}^k u_{i,m} u_{j,n-m}\,.
\end{equation}
It follows that the generating functions $U_k(t) =
\sum_{n=1}^\infty u_{k,n} e^{n \delta t}$ satisfy the same
time evolution equations \rf{Uevol} as before, and that
the couplings $v_{a,n}$ do not modify the leading order result
\rf{Aatrepr} for the correlation functions. However, the
couplings $\hat{v}_{a,n}=v_{a,n}\ep^{n^2\de^2-y_a}$ may diverge
when $\ep\ra 0$. These divergencies cancel the
divergencies which would arise in the
perturbative integrals when removing the cut-off $\ep$.

As pointed out in subsection \ref{TsummingRG2}, it is not clear in general if the
motion  described by the time evolution equations \rf{Uevol} will
remain bounded. It will certainly not remain bounded if the
set $N$ contains relevant fields not equal to the identity
and if some of the corresponding couplings $v_{a,1}$ do not vanish.
The perturbative approach to the construction of
amplitudes in the time-dependent theory will then break down
after a certain time $t$. However, at least in the cases where
the only relevant field contained in $N$ is the identity and where
the couplings $U_k(t)$ determined from the
time evolution equations \rf{Uevol} stay bounded, one may
reliably use our formalism to calculate the
one point functions in the time dependent background.

\section{Aspects of boundary Liouville theory}\label{ABCFT}
\setcounter{equation}{0}

Certain results on
boundary Liouville theory play an important role in our paper.
The relevant results include the
approximate expression
\rf{Fapprox} for the
operator product coefficients $F_{P_\2P_\1}^{P_\3}$
of the boundary fields $\Phi_P^{\de}(x)$ respectively, as well as
the relation \rf{bulkbd} between bulk and boundary fields.
This appendix is devoted to the derivation of these results.

The boundary fields
$\Phi_P^{\de}(x)$ used in this paper are related to the
more general boundary fields $\Psi_\be^{\si_\2\si_1}(x)$
studied in \cite{PT}
by a change of normalization,
\begin{equation}\label{chnorm}
\Phi_P^{\de}(x)\,=\,g^{\si\si}_{\be}
\Psi_\be^{\si\si}(x)\,,\qquad \bigg\{\begin{aligned}
&\be\,=\,1+iP\,,\\[-.5ex] & 2\si\,=\,1-\de\,.
\end{aligned}\bigg\}
\end{equation}
The expression for $g_\be^{\si\si}$ from \cite{PT} may
in the present case ($c=25$, $b=1$) be simplified to
\begin{equation}\label{gfunction}
g_{\be}^{\si\si}\,=\, \mu_r^{\frac{\be}{2}}\,
\frac{\Ga_2(2)\Ga_2(2-2\be)}
     {(\Ga_2(2-\be))^2}
\frac{\Ga_2(2\si)\Ga_2(4-2\si)}
      {\Ga_2(4-2\si-\be)\Ga_2(2\si-\be)}\,.
\end{equation}
The function $\Ga_2(x)$ is known as the Barnes Double Gamma function.
It may be represented by the following integral \cite{FZZ}:
\begin{equation}\label{intrep}
\log\Ga_2(x)\;=\;\int\limits_0^{\infty}\frac{dt}{t}\;
\biggl(\frac{e^{-xt}-e^{-t}}{(1-e^{-t})^2}-
\frac{(1-x)^2}{2e^t}-2\frac{1-x}{t}\biggl)\;.
\end{equation}
It follows from \rf{intrep} that $\Ga_2(x)$ is analytic for ${\rm Re}(x)>0$.

\subsection{Bulk-boundary OPE}\label{BBssect}

Our first aim is to establish the relation \rf{bulkbd} with
$C_\de(P)$ defined by $C_\de(P)\equiv g^{\si\si}_\be$.
This relation is equivalent to the relation
\begin{equation}\label{bulkbd'}
\Psi^{\si\si}_{2\al}(\si)\,|\,B_{\de}^{}\,\lket\,\equiv\,
\lim_{\tau\ra 0} \,(2\tau)^{2\De_\al-\De_{2\al}}\,
V_{\al}(\tau,\si)\,|\,B_{\de}^{}\,\lket\,.
\end{equation}
This relation will be valid
as long as $-1<\de<0$ and ${\rm Re}(\al)<\frac{1}{2}$. It can be
used as a definition of the boundary fields $\Psi^{\si\si}_{\be}(\si)$
for general values of $\si$ and $\be$ thanks to the analyticity
of the fields $\Psi^{\si\si}_{\be}(\si)$ w.r.t. these variables
\cite{T2}.

Our starting point for the derivation of \rf{bulkbd} is the form of the
bulk-boundary OPE valid for
bulk fields $V_{\al}(z,\bz)$ which approach the boundary if $-1<\de<0$ and
$\frac{3}{2}>{\rm Re}(\al)>\frac{1}{2}$,
\begin{equation}\label{BBOPE}
V_{\al}(z,\bz)\,=\,\int_\BS d\be\;|z-\bz|^{\De_\be-2\De_\al}
A^\be_{\al|\de}\,\Psi_{\be}^{\de}(x)+({\rm descendants})\,,
\end{equation}
where $\BS=1+i\BR_+$ and $x={\rm Re}(z)$. The coefficients $A^\be_{\al|\de}$
which appear in the bulk-boundary OPE are related to the expectation value
\begin{equation}\label{BBVEV}
A_{\be|\al}^{\si}\,=\,\lim_{2z\ra {i}}\lim_{x\ra\infty}|x|^{2\De_\be}
\bra\,\Psi^{\si\si}_{\be}(x)\, V_{\al}(z,\bz)\,\ket_{\rm L,\de}^{\BU}
 \end{equation}
via $A^\be_{\de|\al}=A_{Q-\be|\al}^{\si}$, as follows by inserting
\rf{BBOPE} into \rf{BBVEV} and taking into account that
the fields $\Psi^{\si\si}_{\be}(x)$ are normalized by
$\lim_{x\ra\infty}|x|^{2\De_\be}
\bra\,\Psi^{\si\si}_{\be_\2}(x)\Psi^{\si\si}_{\be_\1}(0)\ket_{\rm L,\de}^{\BU}
=\de(\be_\2-\be_\1)$ for $\be_\2,\be_\1\in\BS$.
The explicit expression for $A^{\si}_{\be|\al}$ was found in \cite{Ho}.
It may be represented as
\begin{equation}\label{BBform}\begin{aligned}
& A^{\si}_{\be|\al}\;=\;
\rho_{\be|\al}\int\limits_{-\infty}^{\infty}dt\;
\prod_{\ep=\pm}\frac{S_2\big(\frac{1}{2}(2\al+\be-2)+ i\ep t\big)}
     {S_2\big(\frac{1}{2}(2\al-\be+2)+ i\ep t\big)}\,
e^{4\pi t(\si-1)}\,,\\
&{\rm where}\;\;\rho_{\be|\al}=
\mu^{\frac{2-2\al-\be}{2}}
\frac{\Ga_2^3(2-\be)\Ga_2(4-2\al-\be)\Ga_2(2\al-\be)}
{\Ga_2(2)\Ga_2(\be)\Ga_2(2-2\be)\Ga_2(2\al)\Ga_2(2-2\al)}.
\end{aligned}\end{equation}
Additional discrete terms will appear in the bulk-boundary
OPE \rf{BBOPE} as soon as
${\rm Re}(\al)<\frac{1}{2}$. There is a single discrete term
proportional to $\Psi_{2\al}(x)$ as long
as  $0<{\rm Re}(\al)<\frac{1}{2}$. In order to identify this
contribution let us note that the OPE coefficients
$A^\be_{\de|\al}$ have a pole near $\be=2\al$,
\begin{equation}\label{BBpole}
A^\be_{\de|\al}\,\underset{\be\ra 2\al}{\sim}\,
\frac{1}{2\pi}\,\frac{1}{2\al-\be}\,.
\end{equation}
This is shown by noting that contour of the integration
in \rf{BBform} gets pinched by the two poles of the
integrand at $t=\pm \frac{i}{2}(2\al-\be)$. Extracting the resulting
singular contribution to the integral by contour deformation
yields  \rf{BBpole}.

Without loss of generality\footnote{The other
case leads to the same result thanks to the
symmetry of the integrand in \rf{BBOPE} under $\be\ra 2-\be$.}
let us assume that ${\rm Im}(\al)>0$.
The pole at $\be=2\al$ would cross the contour of integration over $\be$ in
\rf{BBOPE} when varying $\al$ from
${\rm Re}(\al)>\frac{1}{2}$ to
${\rm Re}(\al)<\frac{1}{2}$. The analytic continuation of the
integral in \rf{BBOPE} can be described by integrating over a deformed
contour which is the sum of $\BS$ with a small circle around the
pole at $\be=2\al$. Evaluating the contribution from the latter
by using the residue theorem and \rf{BBpole} leads to the
conclusion that in the range $0<{\rm Re}(\al)<\frac{1}{2}$
the OPE \rf{BBOPE} gets modified to
\begin{equation}\label{BBOPE'}
\begin{aligned}
V_{\al}(z,\bz)\,=\, \int_\BS d\be\;&|z-\bz|^{\De_\be-2\De_\al}
A^\be_{\al|\de}\,\Psi_{\be}^{\de}(x)\\
 +\,& |z-\bz|^{\De_{2\al}-2\De_\al}\,\Psi_{2\al}^{\de}(x)
+({\rm descendants})\,.
\end{aligned}
\end{equation}
The discrete contribution in the second line of \rf{BBOPE'} is the
dominant one for ${\rm Im}(z)\ra 0$. The
sought-for relation \rf{bulkbd'} therefore follows directly from
\rf{BBOPE'}.

Let us remark that the functional equation satisfied
by $\Ga_2(x)$,
\begin{equation} \label{funeq}
\Ga_2(x+1)\,=\,\sqrt{2\pi}\,(\Ga(x))^{-1}\,\Ga_2(x)\,.
\end{equation}
together with the analyticity of $\Ga_2(x)$ for ${\rm Re}(x)>0$ imply
that $\Ga_2(x)$ has a
simple pole at $x=0$.
This implies in particular that
$C_\de(i\de)=g^{\si\si}_{2\si}=0$, as is easily seen from \rf{gfunction}.

It remains to derive the expression \rf{BBd}.
Let us first observe that due to $\bra P|_{\rm L}^{}=
\lim_{z\ra\infty}|z|^{4\De_\al} \bra 0|_{\rm L}^{} V_{Q-\al}(z,\bz)$
for $\al=1+iP$
we have
\begin{equation}\label{bbPhi}
\bra \,P_\1\,|\,\Phi_{P_\2}^{{\de}}(1)\,|\,B_\de^{}\,\ket^{}_{\rm L}\,=\,
g^{\si\si}_{1+iP_\2}\,
A^{\si}_{1+iP_2|1-iP_\1}\,.
\end{equation}
Note furthermore that for $\be=1+iP_2$, $P_\2=\CO(\de)$ we have
$\be-2\si=\de+iP_\2\ra 0$. Formula (B.13) of \cite{T2}
is therefore applicable in the case at hand and gives
\begin{equation}\label{BBapprox}
A^{\si}_{1+iP_2|1-iP_\1}\,
\dasym\,\frac{\mu_r^{\frac{1}{2}}}{2\pi}\,\frac{2P_\2}{P_\2-i\de}\,
\frac{P_\1}{\sinh 2\pi P_\1}\,
\Theta(P)\,.
\end{equation}
Equation \rf{BBd} now follows easily by inserting \rf{BBapprox} and
\rf{gapprox} into \rf{bbPhi}.

\subsection{Structure functions of boundary fields}

It follows from \rf{chnorm} that $F_{P_\2P_\1}^{P_\3}$
and $D_{P_\3P_\2P_\1}$ can be expressed in terms
of the
OPE coefficients $C_{\al_\3|\al_\2\al_\1}^{\si\si\si}$
and
the three point functions $C_{\al_\3\al_\2\al_\1}^{\si\si\si}
\equiv C_{Q-\al_\3|\al_\2\al_\1}^{\si\si\si}$
of the fields $\Psi_\be^{\si_\2\si_1}(x)$ \cite{PT},
\begin{equation}\label{D-C}
D_{P_\3P_\2P_\1}\,=\,g^{\si\si}_{\al_\3}g^{\si\si}_{\al_\2}g^{\si\si}_{\al_\1}
\,C_{\al_\3\al_\2\al_\1}^{\si\si\si}\,,
\qquad
F_{P_\2P_\1}^{P_\3}\,=\,
\frac{g^{\si\si}_{\al_\2}g^{\si\si}_{\al_\1}}{g^{\si\si}_{\al_\3}}
\,C_{\al_\3|\al_\2\al_\1}^{\si\si\si}\,
\end{equation}
The formula for $C_{\al_\3|\al_\2\al_\1}^{\si\si\si}$
from \cite{PT} simplifies  slightly
in the present case,
\begin{align}
{} C_{\alpha_{3}|\alpha_{2}\alpha_{1}}^{\si\si\si}
= &  \mu_r^{\frac{i}{2}(P_\3-P_\2-P_\1)}
\Gamma_2(1-i(P_\1+P_\2+P_\3))\nn\\
& \times \frac{\Gamma_2(1+i(P_\2+P_\3-P_\1))
 \Gamma_2(1+i(P_\2-P_\1-P_\3))\Gamma_2(1+i(P_\3-P_\1-P_\2))}
{\Gamma_2(2iP_\3)\Gamma_2(-2iP_\2)\Gamma_2(-2iP_\1)\Gamma_2(2)}
\nn \\
&   \times
\frac{S_2(1+iP_\3)S_2(2+\de+iP_\3)}
     {S_2(1+iP_\2)S_2(2+\de+iP_\2)}
   \int\limits_{\BR+i0}ds \;\prod_{k=1}^{4}
    \frac{S_2(U_k+is)}
         {S_2(V_k+is)},
\label{Cexp}\end{align}
where we have used the identifications from \rf{chnorm}, the
definitions $S_2(x)=\Ga_2(x)/\Ga_2(2-x)$ and
\[
\begin{aligned}
& U_1 = 
-\de-iP_\1 ,\\
& U_2 = 
1-iP_\1,  \\
& U_3 = 
1-iP_\2, \\
& U_4 = 
1+iP_\2,
\end{aligned}
\qquad  \qquad
\begin{aligned}
& V_1 =
2-i(P_\1-P_\3), \\
& V_2 =
2-i(P_\1+P_\3), \\
& V_3 = 
1-\de, \\
& V_4=2\;.
\end{aligned}
\]
Taking the limit $\de\ra 0$, $P_k=\CO(\de)$ within the formulae
\rf{D-C}, \rf{Cexp} is subtle due to the fact that many
of the factors exhibit singular behavior. First note that
it follows from the functional equation \rf{funeq} that
$\Ga_2(x)$ behaves for $x\ra 0$ as
\begin{equation}
\Ga_2(x)\,\underset{x\ra 0}\sim\,\frac{1}{x}\,\frac{\Ga_2(1)}{\sqrt{2\pi}}\,.
\end{equation}
Let us next consider the integral which appears in
\rf{Cexp}. We observe that the contour of integration
in \rf{Cexp} gets pinched between the poles of the
integrand at $is=-U_\1=\al_\1-2\si=\de(iq_\1-1)$ and $is=2-V_4=0$.
It follows that the integral diverges in this limit. The
singular behavior for $\de\ra 0$ is identified by deforming
the contour of integration into the contour $\BR+i0$ plus a
small circle $\CC_\1$ around the pole at $s=iU_\1$.
The contribution from the
latter is found to be
\begin{align}
\int_{\CC_\1}ds \;\prod_{k=1}^{4}
    \frac{S_2(U_k+is)}
         {S_2(V_k+is)}\,& \dasym\,
{S_2(-\de+iP_\3)S_2(-\de-iP_\3)S_b(-\de-iP_\1)}\\
& \dasym\,\frac{1}{(2\pi)^3}\frac{1}{-\de+iP_\3}\frac{1}{-\de-iP_\3}
\frac{1}{-\de-iP_\1}\,,
\end{align}
where we have used $S_2(x)S_2(2-x)=1$ and $S_2(x)=1$.
Collecting the factors yields
\begin{equation}\label{Capprox}
 C_{\alpha_{3}\alpha_{2}\alpha_{1}}^{\si\si\si} \,\dasym\,
{2\pi}\,
\frac{\mu_r^{-\frac{3}{2}}}{(2\pi)^3}\,\prod_{k=1}^{3}
\frac{2P_k}{P_k+i\de} 
\,.
\end{equation}
One finds similarly
\begin{equation}\label{gapprox}
g_{\be_k}^{\si\si}\,\dasym\, 2\pi\,
\mu_r^{\frac{1}{2}}\,\frac{P_k-i\de}{2P_k}\,\qquad
k=1,2,3\,.
\end{equation}
The approximate
expressions for the three point functions $D_{P_\3P_\2P_\1}$
and operator product coefficients $F_{P_\2P_\1}^{P_\3}$
now follow easily by inserting \rf{gapprox} and \rf{Capprox} into
\rf{D-C}.

\subsection{A note on the normalizations}

The normalizations used in the present paper are fully fixed by the
requirements
\begin{align}
 \lim_{z_\2\ra z_\1}\;& |z_\2-z_\1|^{\2(\De_{\al_\2}+\De_{\al_\1}-
\De_{\al_\2+\al_\1})}\,
V_{\al_\2}(z_\2,\bz_\2)V_{\al_\1}(z_\1,\bz_\1)
\,=\, V_{\al_\1+\al_\2}(z_\1,\bz_\1)\,,\\
\lim_{x_\2\ra x_\1}\;& |x_\2-x_\1|^{\De_{\be_\2}+\De_{\be_\1}-
\De_{\be_\2+\be_\1}}\Psi_{\be_\2}^{\si\si}(x_\2)
\Psi_{\be_\1}^{\si\si}(x_\1)\,=\,
\Psi_{\be_\2+\be_\1}^{\si\si}(x_\1)\,,\label{Bleading}\\
 \lim_{z-\bz \ra 0} \;& |z-\bz|^{2\De_\al-\De_{\2\al}}\,
V_{\al}(z,\bz)\,=\,\Psi_{\2\al}^{\si\si}(x)\,,
\end{align}
which follow from
\begin{align}
&V_{\al}(z,\bz)\,\underset{\phi_0\ra-\infty}{\sim}\,
:\!e^{2\al\phi(z,\bz)}\!: \,,\\
&\Psi_{\be}^{\si\si}(x)\,\underset{\phi_0\ra-\infty}{\sim}\,
:\!e^{\be\phi(x)}\!: \,.
\end{align}
In order to show that \rf{Cexp} implies \rf{Bleading} one may proceed
as in the discussion of the bulk-boundary OPE in subsection
\ref{BBssect}. Equation \rf{Bleading} follows from the observation that
\begin{equation}
C_{\alpha_{3}|\alpha_{2}\alpha_{1}}^{\si\si\si} \,
\underset{\al_3\ra \al_1+\al_2}{\sim}\,\frac{1}{2\pi}\,
\frac{1}{\al_1+\al_2-\al_3}\,,
\end{equation}
which is shown by the same method as \rf{BBpole}.
The normalization prescriptions used in \cite{FZZ,Ho,T2}
differ slightly from the one used in this paper.


\begin{thebibliography}{77}

\bibitem[FHL]{FHL}
D.Z. Freedman, M. Headrick, A. Lawrence,
``On Closed String Tachyon Dynamics'',
Phys.Rev. {\bf D73} (2006) 066015
[arXiv:hep-th/0510126]

\bibitem[GM]{GM}  P. Ginsparg, G. Moore,
  ``Lectures on 2D gravity and 2D string theory (TASI 1992)'',
  in {\it Recent Directions in Particle Theory}, eds. J. Harvey and
  J. Polchinski, Proceedings of the 1992 TASI, World Scientific,
  Singapore, 1993

\bibitem[SS]{SS} N. Seiberg, D. Shih,
``Minimal string theory'',
Comptes Rendus Physique {\bf 6} (2005) 165-174
[arXiv:hep-th/0409306]

\bibitem[CL]{CL} A. Cappelli, J.I. Latorre,
``Perturbation theory of higher-spin conserved currents off criticality'',
Nucl. Phys. {\bf B340} (1990) 659-691



\bibitem[FZZ]{FZZ}
V.~Fateev, A.~B.~Zamolodchikov and A.~B.~Zamolodchikov,
``Boundary Liouville field theory. I: Boundary state and boundary  two-point function,'' [arXiv:hep-th/0001012].

\bibitem[T0]{TL}
J.~Teschner,
``Liouville theory revisited,''
Class.\ Quant.\ Grav.\  {\bf 18}, R153 (2001)
[arXiv:hep-th/0104158].


\bibitem[T1]{T1}
J. Teschner,
``Remarks on Liouville with a boundary'', Talk at TMR-conference "Nonperturbative Quantum Effects 2000
[arXiv:hep-th/0009138].

\bibitem[Ho]{Ho} K. Hosomichi,
``Bulk-Boundary Propagator in Liouville Theory on a Disc'',
JHEP {\bf 0111} (2001) 044

\bibitem[ZZ]{ZZ} A.~B.~Zamolodchikov and A.~B.~Zamolodchikov,
``Structure constants and conformal bootstrap in Liouville field theory,''
Nucl.\ Phys.\ B {\bf 477}, 577 (1996)
[arXiv:hep-th/9506136].

\bibitem[PT]{PT}
B.~Ponsot and J.~Teschner,
``Boundary Liouville field theory: Boundary three point function,''
Nucl.\ Phys.\ B {\bf 622}, 309 (2002)
[arXiv:hep-th/0110244].

\bibitem[T2]{T2} J. Teschner,
``On boundary perturbations in Liouville theory and brane
dynamics in noncritical string theory'',
JHEP\ 04 (2004) 023

\bibitem[DIR]{DIR} A. Dabholkar, A. Iqubal, J. Raeymaekers,
``Off-Shell Interactions for Closed-String Tachyons'',
JHEP {\bf 0405} (2004) 051 [arXiv:hep-th/0403238]

\bibitem[SarSat]{SarSat} S. Sarkar, B. Sathiapalan,
``Closed String Tachyons on $C/Z_N$'', Int.J.Mod.Phys. {\bf A19} (2004) 5625-5638
[arXiv:hep-th/0309029]

\bibitem[SeibSh]{SeibSh}  N. Seiberg, S. Shenker,
``A Note on Background (In)dependence'', Phys.Rev. {\bf D45} (1992) 4581-4587
[arXiv:hep-th/9201017]


\bibitem[APS]{APS}  A. Adams, J. Polchinski, E. Silverstein,
``Don't Panic! Closed String Tachyons in ALE Spacetimes'',
JHEP {\bf 0110} (2001) 029
[arXiv:hep-th/0108075]




\end{thebibliography}
\end{document}